\documentclass[submission,copyright,creativecommons]{eptcs}
\providecommand{\event}{ICE 2023} 

\usepackage{float}
\usepackage{tikz}
\usetikzlibrary{automata,arrows,positioning,shapes.geometric}
\tikzset{elliptic state/.style={draw,ellipse}}
\tikzstyle{mybox} = [draw=black, fill=white, very thick, rectangle, rounded corners, inner sep=8pt, inner ysep=8pt]
\tikzstyle{fancytitle} = [fill=white, text=black, very thick, align=center, rounded corners]
\usepackage{iftex}

\ifpdf
  \usepackage{underscore}         
  \usepackage[T1]{fontenc}        
\else
  \usepackage{breakurl}           
\fi

\title{Research Challenges in Orchestration Synthesis}
\author{Davide Basile \qquad\quad Maurice H. ter Beek
\institute{Formal Methods and Tools lab, ISTI--CNR, Pisa, Italy}
\email{\{davide.basile,maurice.terbeek\}@isti.cnr.it}
}
\def\titlerunning{Research Challenges in Orchestration Synthesis}
\def\authorrunning{Davide Basile and Maurice H. ter Beek}

\usepackage{amsfonts,amsmath,amstext,latexsym,MnSymbol,amsthm}
\usepackage[misc]{ifsym}
\usepackage{xcolor}
\usepackage{graphicx}
\usepackage{booktabs,colortbl}
\usepackage{wrapfig}
\usepackage{multirow}

\usepackage{doi,url}

\usepackage{xifthen}
\newcommand{\inv}[1][]{%
\ifthenelse{\equal{#1}{}}{I}{I(#1)}%
}
\newcommand{\con}[1][]{%
\ifthenelse{\equal{#1}{}}{\chi}{\chi(#1)}%
}

\newcommand{\Permitted}{{\scalebox{1.1}{$\circ$}}}

\newcommand{\offerset}{A^o}

\newcommand{\ithel}[2]{{#1}_{({#2})}}

\newcommand{\blk}{\bullet}
\newcommand{\Oset}{\mathsf{O}}
\newcommand{\Rset}{\mathsf{R}}

\newcommand{\TRANS}[1]{\xrightarrow{#1}}

\newcommand{\TRANSS}[1]{\raisebox{-.3ex}[0pt][0pt]{$\xrightarrow{\raisebox{-.3ex}[0pt][0pt]{\scriptsize $#1$}}$}}

\renewcommand{\epsilon}{\varepsilon}

\renewcommand{\emptyset}{\diameter}

\newcommand{\hbra}{
\hbox to 1 \textwidth{\vrule width0.3mm height 1.8mm depth-0.3mm
                    \leaders\hrule height1.8mm depth-1.5mm\hfill
                    \vrule width0.3mm height 1.8mm depth-0.3mm}}
\newcommand{\hket}{
\hbox to 1 \textwidth{\vrule width0.3mm height1.5mm
                    \leaders\hrule height0.3mm\hfill
                    \vrule width0.3mm height1.5mm}}

\makeatletter
\def\thickhline{%
  \noalign{\ifnum0=`}\fi\hrule \@height \thickarrayrulewidth \futurelet
   \reserved@a\@xthickhline}
\def\@xthickhline{\ifx\reserved@a\thickhline
               \vskip\doublerulesep
               \vskip-\thickarrayrulewidth
             \fi
      \ifnum0=`{\fi}}
\makeatother
\newlength{\thickarrayrulewidth}
\usepackage[most]{tcolorbox}
\newtcolorbox{markbox}{
  enhanced,
  breakable,
  size=minimal,
  parbox=false,
  after={\par},
  before upper={\indent},
  colback=white,
  overlay = {
	 \draw[line width=2pt]
	 (frame.north east) -| ([xshift=3mm]frame.east) |-(frame.south east);
  },
  overlay first={\draw[line width=2pt] (frame.north east) -| ([xshift=3mm]frame.south east);},
  overlay middle={\draw[line width=2pt] ([xshift=3mm]frame.north east) -- ([xshift=3mm]frame.south east);},
  overlay last={\draw[line width=2pt] ([xshift=3mm]frame.north east) |- (frame.south east);},
}

\usepackage{wrapfig}

\usepackage{lipsum}  

\newtheorem{definition}{Definition}
{\theoremstyle{definition}\newtheorem{example}{Example}}

\begin{document}

\maketitle
 
\begin{abstract}
Contract automata allow to formally define the behaviour of service contracts in terms of service offers and requests, some of which are moreover optional and some of which are necessary. A composition of contracts is said to be in agreement if all service requests are matched by corresponding offers. Whenever a composition of contracts is not in agreement, it can be refined to reach an agreement using the orchestration synthesis algorithm. This algorithm is a variant of the synthesis algorithm used in supervisory control theory and it is based on the fact that optional transitions are controllable, whereas necessary transitions are at most semi-controllable and cannot always be controlled. In fact, the resulting orchestration is such that as much of the behaviour in agreement is maintained. 
In this paper, we discuss recent developments of the orchestration synthesis algorithm for contract automata. Notably, we present a refined notion of semi-controllability  and compare it with the original notion by means of examples. We then discuss the current limits of the orchestration synthesis algorithm and identify a number of  research challenges together with a research roadmap.
\end{abstract}

\section{Introduction}

Orchestrations of services describe how control and data exchanges are coordinated in distributed service-based applications and systems.
Their principled design is identified in~\cite{manifesto} as one of the primary research challenges for the next $10$~years, 
and the Service Computing Manifesto~\cite{manifesto} points out that \lq\lq Service systems have so far been built without an adequate rigorous foundation that would enable reasoning about them\rq\rq\ and, moreover, that \lq\lq The design of service systems should build upon a formal model of services\rq\rq.

The problem of synthesising  well-behaving orchestrations of services can be viewed as a specific instance of the more general problem of synthesising strategies in games~\cite{BBP20,BBBC23}. 
This can be solved using refined algorithms from supervisory control for discrete event systems~\cite{RW87,AMPS98}, which have well-established relationships with reactive systems synthesis~\cite{DBLP:journals/deds/EhlersLTV17}, parity games~\cite{DBLP:journals/acta/LuttenbergerMS20}, automated behaviour composition~\cite{7473906} and automated planning~\cite{DBLP:conf/aips/CamachoBM19}.

Contract automata are a specific type of finite state automata that are used to formally define the behaviour of service contracts. These automata express contracts in terms of both offers and requests~\cite{BDF16}.
When multiple contracts are composed, they are said to be in agreement if all service requests from one contract are matched by another contract's corresponding offers. A composition of contracts that is not in agreement, can automatically be refined to reach an agreement by means of the orchestration synthesis algorithm, which is a variation of the synthesis algorithm used in supervisory control theory. 
This orchestration synthesis algorithm for contract automata is described in~\cite{BBDLFGD20,BBP20}.

The classic algorithm for synthesising a most permissive controller distinguishes transitions whose controllability is invariant~\cite{RW87,AMPS98}. 
In service contracts, instead, the controllability of certain transitions may vary depending on specific conditions on the orchestration of contracts~\cite{BBP20}.
The contract automata library {\tt CATLib}~\cite{BB22OSP} implements contract automata and their operations (e.g., composition and synthesis). 
%
%
Orchestrations of contract automata abstract from their underlying realisation; an orchestrator is assumed to interact with the services to realise the overall behaviour as prescribed by the orchestration contract. 
The contract automata runtime environment {\tt CARE}~\cite{BB23} implements an orchestrator that interprets the synthesised orchestration to coordinate the services, where each service is implementing a contract. Thus, {\tt CARE} is explicating the low-level interactions that are abstracted in contract automata orchestrations. 
Notably, one aspect that is abstracted in contract automata and concretised at the implementation level is that of selecting the next transition to execute in the presence of choice. In~\cite{BB23}, different implementations are proposed based on whether services may participate externally or internally in a choice.

This paper delves into challenges and research issues for orchestration synthesis of contract automata, given the latest developments in this field. In particular, we start by refining the current definition of semi-controllability to consider the aforementioned possible realisations of choices defined in~\cite{BB23}. We provide several examples to illustrate the differences between the refined definition and the original definition. 
The various definitions of semi-controllability lead to different sets of contract automata orchestrations, which we present in Figure~\ref{fig:venndiagram} together with an example for each level of the orchestration hierarchy depicted. This allows us to highlight the unique characteristics of each level and to identify current issues in synthesising orchestrations of contract automata using these examples. 
Based on the issues presented, we then outline future research challenges in the orchestration synthesis of contract automata and a research roadmap to address them.

\paragraph{Related Work}
At last year's ICE 2022 workshop, the compositionality of communicating finite state machines (CFSM) with asynchronous semantics was discussed in~\cite{DBLP:journals/corr/abs-2208-04634}. 
Also contract automata are composable, enabling the modelling of systems of systems. 
Moreover, under certain specific conditions that were presented at the 2014 edition of ICE~\cite{DBLP:journals/corr/BasileDFT14,BDFT15}, an orchestration of contract automata can be translated into a choreography of synchronous or asynchronous CFSM. 
The relation between  multiparty session types and CFSM is discussed in~\cite{DBLP:conf/fct/YoshidaZF21}. 
Therefore, contract automata can be related to  multiparty session types by exploiting their common relation with CFSM~\cite{DBLP:journals/corr/BasileDFT14,BDFT15,DBLP:conf/fct/YoshidaZF21}.

The contract automata approach is closer to~\cite{DBLP:journals/scp/KouzapasDPG18}, in which behavioural types are expressed as finite state automata of {\tt Mungo}, called typestates~\cite{DBLP:journals/tse/StromY86}.
Similarly to {\tt CARE}, the runtime environment for contract automata~\cite{BB23}, in {\tt Mungo} finite state automata are used as behaviour assigned to Java classes (one automaton per class), with transition labels corresponding to methods of the classes.
A tool to translate typestates into automata was presented at ICE 2020~\cite{DBLP:journals/corr/abs-2009-08769}. {\tt CATApp}, a graphical front-end tool for designing contract automata, is available in~\cite{CATAPPurl}.
A tool similar to {\tt Mungo} is {\tt JaTyC} (Java Typestate Checker)~\cite{BACCHIANI2022102844}. 

The refined definition of semi-controllability presented in this paper closely aligns with the notion of weak receptiveness in team automata~\cite{BCHK17,BCHP23}. However, the challenges addressed in this paper are primarily related to the problem of synthesising an orchestration of services and as such are not directly relevant to team automata.

Differently from the semi-controllability for  orchestrations, a distinct notion of semi-controllability has been studied in~\cite{BBP20,BasileB21} for choreographies of services. Finally, while a runtime environment for the orchestration of services has been proposed in~\cite{BB23}, this has yet to be realised for the case of choreographies, which could result in improvements in the notion of semi-controllability for choreographies.

\paragraph{Outline} 
We start by providing some background on contract automata and orchestration synthesis in Section~\ref{sect:background}.
We introduce a refined notion of semi-controllability in Section~\ref{sect:semicontrollability}. 
In Section~\ref{sect:challenges}, we present several research challenges for orchestration synthesis of contract automata. 
We conclude in Section~\ref{sect:conclusion}.

\section{Background}\label{sect:background}

We will begin by formally introducing contract automata and their synthesis operation. Contract automata are a type of finite state automata that use a partitioned alphabet of actions. A Contract Automaton (CA) can model either a single service or a composition of multiple services that perform actions. The number of services in a CA is known as its rank. If the rank of a CA is 1, then the contract is referred to as a principal (i.e., a single service).

The labels of a CA are vectors of atomic elements known as actions. Actions are categorised as either requests (prefixed by $?$), offers (prefixed by $!$), or idle actions (represented by a distinguished symbol $-$). Requests and offers belong to the sets $\Rset$ and $\Oset$, respectively, and they are pairwise disjoint. The states of a CA are vectors of atomic elements known as basic states. 
Labels are restricted to requests, offers or matches. In a request (resp. offer) label there is a single request (resp. offer) action and all other actions are idle. In a match label there is a single pair of request and offer actions that match, and all other actions are idle. The length of the vectors of states and labels is equal to the rank of the CA.
For example, the label $[!a,?a,-,-]$ is a match where the request action $?a$ is matched by the offer action $!a$, and all other actions are idle.  
Note the difference between a request label (e.g., $[?a,-]$) and a request action (e.g., $?a$). 
A transition may also be called a request, offer or match according to its label. 
Figure~\ref{fig:principalsABC} depicts three principal contracts, whilst Figure~\ref{fig:orchestrationABC} depicts a contract of rank~3.

The goal of each service is to reach an accepting (\emph{final}) state such that all its request (and possibly offer) actions are matched.
Transitions are equipped with \emph{modalities}, i.e., \emph{necessary\/}~($\Box$) and \emph{optional\/}~($\Permitted$) transitions, respectively\,\footnote{Originally, in~\cite{BBDLFGD20}, the optional modality was called permitted and denoted with $\Diamond$. Since in contract automata the two modalities are a partition, the terminology has been updated to avoid confusion with modal transition systems, where $\Box \subseteq \Diamond$.}.
Optional transitions are controllable, whereas necessary transitions can be uncontrollable (called \emph{urgent} necessary transitions) or semi-controllable (called \emph{lazy}  necessary transitions). 
The resulting formalism is called \emph{Modal Service Contract Automata} (MSCA). 
In the following definition, given a vector $\vec a$, its $i$th element is denoted by $\ithel {\vec a} i$.

\begin{definition}[MSCA]\label{def:contract}
Given a finite set of states $\mathcal{Q} = \{q_1,q_2, \ldots \}$, an 
MSCA $\mathcal{A}$ of rank $n$ is a tuple $\langle Q, \vec{q_0}, A^r, A^o, T, F \rangle$, 
with set of states $Q=Q_1 \times \ldots \times Q_n \subseteq \mathcal{Q}^n$, 
initial state $\vec{q_0} \in Q$,
set of requests $A^{r} \subseteq \Rset$,
set of offers $A^{o} \subseteq \Oset$, set of final states $F \subseteq Q$,
set of transitions $T \subseteq Q \times A  \times Q$, where 
$A\subseteq(A^r \cup \offerset \cup \{\blk\})^n$,  
partitioned into \emph{optional\/} transitions $T^\Permitted$ and \emph{necessary\/} transitions $T^\Box$, with $T^\Box$ further partitioned into \emph{urgent\/} necessary transitions $T^{\Box_u}$ and \emph{lazy\/} necessary transitions $T^{\Box_l}$, and such that given $t= (\vec{q},\vec{a},\vec{q}\,') \in T$: 
i)~$\vec{a}$ is either a request, an offer or a match;
ii)~if $\vec a$ is an offer, then $t \in T^\Permitted$; 
and iii)~$\forall i \in 1\ldots n,\ \ithel{\vec{a}} i=\blk$ implies $\ithel{\vec{q}} i=\ithel{\vec{q}'} i$.
\end{definition}

Composition of services is rendered through the composition of their MSCA models by means of the \emph{composition operator\/} $\otimes$, which is a variant  of a synchronous product. 
This operator basically interleaves or matches the transitions of the component MSCA, but, whenever two component MSCA are enabled to execute their respective request/offer action, then the match is forced to happen.
Moreover, a match involving a necessary transition of an operand is itself necessary.
The rank of the composed MSCA is the sum of the ranks of its operands. The vectors of states and actions of the composed MSCA  are built from the vectors of states and actions of the component MSCA, respectively.
In this paper, we will only consider principal contracts and compositions of principals, which will be automatically refined into orchestrations (as shown in Figure~\ref{fig:orchestration}). However, it is important to note that contracts can be created by composing contracts with a rank of one or higher.

In a composition of MSCA, typically various properties are analysed. We are especially interested in \emph{agreement\/}.
The property of agreement requires to match all requests, whereas offers can go unmatched.

CA support the synthesis of the most permissive controller (mpc) known from the theory of supervisory control of discrete event systems~\cite{RW87,CL06}, where a finite state automaton model of a \emph{supervisory controller\/} is synthesised from given (component) finite state automata that are composed. 
The synthesised automaton, if successfully generated (i.e., non-empty), is such that it is \emph{non-blocking}, \emph{controllable}, and \emph{maximally permissive}. 
An automaton is said to be \emph{non-blocking\/} if, from each state, at least one of the \emph{final states\/} (distinguished stable states that represent completed \lq tasks\rq~\cite{RW87}) can be reached without passing through so-called \emph{forbidden states}, meaning that there is always a possibility to return to an accepted stable state (e.g., a final state). 

The synthesised automaton is said to be \emph{controllable} when only controllable transitions are disabled. Indeed, the supervisory controller is not permitted to directly block uncontrollable transitions from occurring; the controller is only allowed to disable them by preventing controllable actions from occurring.\linebreak 
Finally, the fact that the resulting supervisory controller is said to be \emph{maximally permissive\/} (or least restrictive) means that as much behaviour of the uncontrolled system as possible is present in the controlled system without violating neither the requirements, nor controllability nor the non-blocking condition.

\paragraph{Orchestration Synthesis}

As stated previously, optional transitions are controllable, whereas necessary transitions can be either uncontrollable (called {\it urgent}) or semi-controllable (called {\it lazy}).  
In the mpc synthesis (implemented in {\tt CATLib}~\cite{BBP20,BB22OSP}), all necessary transitions are {\it urgent}, i.e., they are always uncontrollable. 
This stems from the fact that traditionally uncontrollable transitions relate to an unpredictable environment.

When synthesising an orchestration of services, all necessary transitions are instead {\it lazy}, i.e., they are  \emph{semi-controllable\/}~\cite{BBDLFGD20,BBP20}.
A semi-controllable transition $t$ is a transition that is either uncontrollable or controllable according to given conditions.  
In~\cite{BBP20}, different conditions are given according to whether the synthesis of an orchestration or a choreography is computed. 
In this paper, we only consider orchestrations. 
Below, we denote with $\textit{Dangling\/}(\mathcal A)$ the set of states that are not reachable from the initial state or cannot reach any final state. 
More in detail, a semi-controllable transition $t$ is controllable if in a given portion $\mathcal A'$ of $\mathcal{A}$ there exists a semi-controllable match transition $t'$, with source and target states not dangling,  such that in both $t$ and $t'$ the \emph{same\/} service, in the \emph{same\/} local state, does the \emph{same\/} request. Otherwise, $t$ is uncontrollable. 

\begin{definition}[Controllability]\label{def:controllabilityorchestration}
Let $\mathcal A$ be an MSCA and let $t = (\vec q_1, \vec a_1, \vec q_1\!') \in T_{\mathcal A}$. Then:
\begin{itemize}
\item if $t \in T^\Permitted_{\mathcal A}$, then  $t$ is \emph{controllable\/} (in $\mathcal A$);
\item if $t \in T^{\Box_u}_{\mathcal A}$, then  $t$ is \emph{uncontrollable\/} (in $\mathcal A$);
\item if $t \in T^{\Box_l}_{\mathcal A}$, then $t$ is \emph{semi-controllable\/} (in $\mathcal A$).
\end{itemize}
Moreover, given $\mathcal A'\subseteq\mathcal A$, if $t$ is semi-controllable and 
$\exists\,t' = (\vec{q_2}, \vec {a_2}, \vec{q_2}\!') \in T^\Box_{{\mathcal A}'}$ in $\mathcal A'$ such that $\vec a_2$ is a match,  
$\vec q_2, \vec q_2\!' \not\in \textit{Dangling\/}(\mathcal A')$,
$\ithel{\vec{q}_1{}}{i} = \ithel{\vec{q}_2{}}{i}$,  and $\ithel{\vec{a}_1{}}{i} = \ithel{\vec{a}_2{}}{i} = ?a$  for some $i \in 0 \ldots rank(\mathcal A)$, 
then $t$ is \emph{controllable\/} in $\mathcal A'$ (via $t'$).
Otherwise, $t$ is \emph{uncontrollable\/} in $\mathcal A'$.
\end{definition}

The interpretation of optional/controllable and urgent/uncontrollable transitions is standard~\cite{RW87,CL06}.
In the upcoming section, we will delve into different understandings and interpretations of the concept of semi-controllability.  
We remark that the orchestration synthesis defined below does not support urgent transitions. 
The orchestration synthesis, as defined below, involves an iterative refinement of the initial automaton $\mathcal A$ (i.e., the composition of contracts). In each iteration, transitions are selectively pruned, and a set $R$ of forbidden states is updated accordingly.
A transition $t$ is pruned under one of two conditions: if it is a request (thus violating the agreement property enforced by the orchestration), or if the target state of $t$ belongs to the set $R$ computed up to that point.
During the first iteration, all request transitions, including both  lazy and optional ones, are pruned. 

In Definition~\ref{def:controllabilityorchestration}, the automaton $\mathcal A'$  represents an intermediate refinement of $\mathcal A$ (the starting composition) which occurs during an iteration  of the synthesis process.
Intuitively, the semi-controllable transition $t$ of $\mathcal A$ is controllable in $\mathcal A'$ because there is another transition $t'$ in $\mathcal A'$ matching the same request from the same service in the same state. 
Otherwise, if there is no such transition $t'$ in $\mathcal A'$, then $t$ is uncontrollable. 
 Put differently, the controllability of $t$ in $\mathcal A'$ relies on the presence of a corresponding transition $t'$ within $\mathcal A'$ itself. If such a matching transition $t'$ does not exist in $\mathcal A'$, then $t$ is deemed uncontrollable.

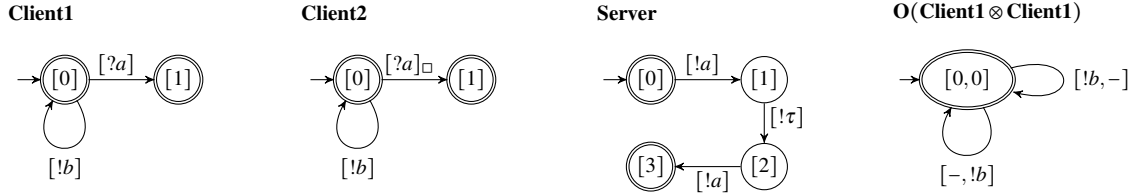
\begin{figure}[t]
\begin{center}
  \begin{tikzpicture}[>=stealth', every state/.style={draw, minimum 
      size=15pt, inner sep=1.5pt}, scale=0.90, node distance=25pt] 
    \node[state,double] (0) {\scriptsize $[0]$};
    \node[state,double,right=of 0] (1) {\scriptsize $[1]$};
    \path ([xshift=-10pt,yshift=28pt] 0) node {\scriptsize\bf Client1}; 
    \draw[->] 
    (0)++(-0.7,0) -- (0)
    (0) edge [loop below,out=300,in=240,looseness=8] node[below] {\scriptsize $[!b]$} (0)   
    (0) edge node[xshift=-2pt,yshift=-2pt,above] {\scriptsize $[?a]$} (1); 

    \node[state,double,right=50pt of 1] (0bis) {\scriptsize $[0]$};
    \node[state,double,right=of 0bis] (1bis) {\scriptsize $[1]$};
    \path ([xshift=-10pt,yshift=28pt] 0bis) node {\scriptsize\bf Client2}; 
    \draw[->] 
    (0bis)++(-0.7,0) -- (0bis)
    (0bis) edge [loop below,out=300,in=240,looseness=8] node[below] {\scriptsize $[!b]$} (0bis)   
    (0bis) edge node[xshift=-2pt,yshift=-2pt,above] {\scriptsize $[?a]_\Box$} (1bis);     

    \node[state,double,right=50pt of 1bis] (s0) {\scriptsize $[0]$};
    \node[state,right=of s0] (s1) {\scriptsize $[1]$};
    \node[state,below=15ptof s1] (s2) {\scriptsize $[2]$};
    \node[state,double,left=of s2] (s3) {\scriptsize $[3]$};
    \path ([xshift=-10pt,yshift=28pt] s0) node {\scriptsize\bf Server}; 
    \draw[->] 
    (s0)++(-0.7,0) -- (s0)
    (s0) edge node[xshift=-2pt,yshift=-2pt,above] {\scriptsize $[!a]$} (s1)   
    (s1) edge node[xshift=-2pt,yshift=2pt,right] {\scriptsize $[!\tau]$} (s2)   
    (s2) edge node[xshift=2pt,yshift=2pt,below] {\scriptsize $[!a]$} (s3);
      
    \node[elliptic state,double,right=50pt of s1] (cc) {\scriptsize $[0,0]$};
    \path ([xshift=3.3cm,yshift=28pt] s1) node {\scriptsize $\text{\bf O}(\text{\bf Client1}\otimes\text{\bf Client1})$};
    \draw[->] 
    (cc)++(-1,0) -- (cc)
    (cc) edge [loop below,out=300,in=240,looseness=6] node[below] {\scriptsize $[-,!b]$} (cc)   
    (cc) edge [loop right] node[right] {\scriptsize $[!b,-]$} (cc);
  \end{tikzpicture} 
\end{center}
\vspace*{-.5\baselineskip}
\caption{\label{fig:principals}Contracts of {\bf Client1}, {\bf Client2} and {\bf Server}, and orchestration $\text{\bf O}(\text{\bf Client1}\otimes\text{\bf Client1})$}
\end{figure}
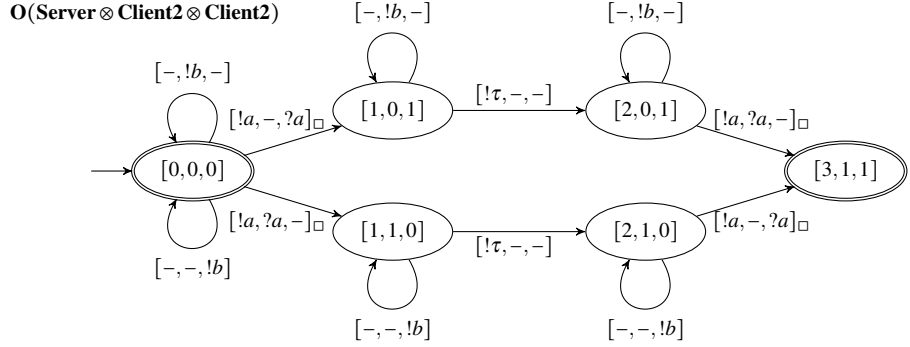
\begin{figure}
\begin{center}
  \begin{tikzpicture}[>=stealth', every state/.style={draw, minimum 
      size=15pt, inner sep=1.5pt}, scale=0.90, node distance=50pt] 
    \node[elliptic state,double] (a0) {\scriptsize $[0,0,0]$};
    \node[elliptic state,above right=of a0,xshift=8pt,yshift=-28pt] (a1) {\scriptsize $[1,0,1]$};
    \node[elliptic state,right=of a1] (a3) {\scriptsize $[2,0,1]$};
    \node[elliptic state,double,below right=of a3,xshift=8pt,yshift=28pt] (a5) {\scriptsize $[3,1,1]$};
    \node[elliptic state,below right=of a0,xshift=8pt,yshift=28pt] (a2) {\scriptsize $[1,1,0]$};
    \node[elliptic state,right=of a2] (a4) {\scriptsize $[2,1,0]$};
    \path ([xshift=-20pt,yshift=66pt] a0) node {\scriptsize $\text{\bf O}(\text{\bf Server}\otimes\text{\bf Client2}\otimes\text{\bf Client2})$}; 
    \draw[->] 
    (a0)++(-1.5,0) -- (a0)
    (a0) edge node[xshift=-6pt,above] {\scriptsize $[!a,-,?a]_\Box$} (a1)   
    (a0) edge [loop above,out=-300,in=-240,looseness=6] node[above] {\scriptsize $[-,!b,-]$} (a0)
    (a0) edge [loop below,out=300,in=240,looseness=6] node[below] {\scriptsize $[-,-,!b]$} (a0)
    (a1) edge node[xshift=-2pt,yshift=-2pt,above] {\scriptsize $[!\tau,-,-]$} (a3)
    (a1) edge [loop above,out=-300,in=-240,looseness=6] node[above] {\scriptsize $[-,!b,-]$} (a1)
    (a3) edge node[xshift=6pt,above] {\scriptsize $[!a,?a,-]_\Box$} (a5)
    (a3) edge [loop above,out=-300,in=-240,looseness=6] node[above] {\scriptsize $[-,!b,-]$} (a3)
    (a0) edge node[xshift=-6pt,below] {\scriptsize $[!a,?a,-]_\Box$} (a2)   
    (a2) edge node[xshift=-2pt,yshift=2pt,below] {\scriptsize $[!\tau,-,-]$} (a4)
    (a2) edge [loop below,out=300,in=240,looseness=6] node[below] {\scriptsize $[-,-,!b]$} (a2)
    (a4) edge node[xshift=6pt,below] {\scriptsize $[!a,-,?a]_\Box$} (a5)
    (a4) edge [loop below,out=300,in=240,looseness=6] node[below] {\scriptsize $[-,-,!b]$} (a4);
  \end{tikzpicture} 
\end{center}
\vspace*{-.5\baselineskip}
\caption{\label{fig:orchestration}Orchestration $\text{\bf O}(\text{\bf Server}\otimes\text{\bf Client2}\otimes\text{\bf Client2})$}
\end{figure}

 Note that in Definition~\ref{def:controllabilityorchestration}, it is not required for $t$ and $t'$ to be distinct. This implies that during the synthesis process, a semi-controllable match transition $t$ can switch from being controllable to uncontrollable only after it has been pruned in a previous iteration.
 To clarify further, a semi-controllable match transition $t$ can switch its controllability status from controllable to uncontrollable only when $t$ is absent in the sub-automaton $\mathcal A'$ during the current iteration. If $t$ is present in $\mathcal A'$ (i.e., it has not been pruned thus far), then, according to Definition~\ref{def:controllabilityorchestration}, $t$ is considered semi-controllable and controllable within $\mathcal A'$ via $t$ itself.
It is important to note that these considerations are applicable only if $t$ is a match. Additionally, it is never the case that a semi-controllable transition $t$ switches from uncontrollable to controllable since transitions are only removed during the synthesis process and are never added back.

The set $R$ of forbidden states is updated at each iteration by adding source states of uncontrollable transitions and dangling states of the refined automaton in the current iteration.  
 Specifically, when the synthesis process eliminates all transitions $t'$ that satisfy the conditions for rendering the semi-controllable transition $t$ controllable via $t'$, then $t$ becomes uncontrollable within the sub-automaton in the current iteration. 
 It is worth noting that even if $t$ was previously pruned in an earlier iteration, its source state $\vec q_1$ might still be reachable in the sub-automaton of the current iteration. Consequently, $\vec q_1$ is added to the set $R$. 
 In the subsequent iteration, all transitions with target state $\vec q_1$ will be pruned. 
 This pruning of transitions whose target is $\vec q_1$ can potentially render another previously pruned semi-controllable transition as uncontrollable, thereby adding its source state to the updated set $R$. 
 This refinement process continues until no further transitions are pruned, and no additional states are added to $R$. The resulting refined automaton obtained at the end of the synthesis process represents the orchestration automaton.

The algorithm for synthesising an orchestration enforcing agreement of MSCA is defined below. 

\begin{definition}[MSCA orchestration synthesis]\label{def:synthesisorchestration}
Let $\mathcal{A}$ be an MSCA and let $\mathcal{K}_{0} = \mathcal A$ and $R_{0} = \textit{Dangling\/}(\mathcal{K}_{0})$.
We let the \emph{orchestration synthesis function\/} $f_o: \textit{MSCA} \times 2^Q \rightarrow \textit{MSCA} \times 2^Q$ be defined as follows:
\begin{center}
\def\arraystretch{1.2}\begin{tabular}{c@{\hskip 0.5in}r@{\hskip 0.05in}c@{\hskip 0.05in}l}
\multicolumn{4}{l}{
$f_o(\mathcal{K}_{i-1},R_{i-1}) = (\mathcal{K}_{i},R_{i}), 
\text{ with }$}\\
& $T_{\mathcal{K}_{i}}$ & = & $T_{\mathcal{K}_{i-1}}\setminus 
\{\,(\vec{q} \TRANS{}\vec{q}\,') = t\in T_{\mathcal{K}_{i-1}} \mid (\vec{q}\,'\!\in R_{i-1}\, \vee t \textit{ is a request})\}$
\\
& $R_{i}$ & = & $R_{i-1} \cup
\{\,\vec q \mid (\vec q \TRANS{})\in T^{\Box_l}_{\mathcal A} \textit{ is uncontrollable in } \mathcal{K}_{i}\,\} \cup \textit{Dangling\/}(\mathcal{K}_{i})$
\end{tabular}
\end{center}
\end{definition}

The orchestration automaton is obtained from the fixpoint of the function $f_o$.
In the rest of the paper, if not stated otherwise, all necessary transitions in the examples are lazy (cf.\ Definition~\ref{def:contract}); for brevity and less cluttering in the figures, we denote them by $\Box$ rather than $\Box_l$.

\begin{example}\label{ex:clients}
We provide an illustrative example to underline the differences between optional transitions, urgent necessary transitions and lazy necessary transitions. 
Figure~\ref{fig:principals} shows two client contracts and a server contract. 
Firstly, we discuss the difference between optional and necessary transitions. 
When all actions of the client contract are optional ({\bf Client1}), there exists an orchestration of the composition of two {\bf Client1} contracts, also depicted in Figure~\ref{fig:principals} ($\text{\bf O}(\text{\bf Client1}\otimes\text{\bf Client1})$). 
Indeed the (transition labelled with the) request~$?a$ is optional and can be removed to obtain the orchestration. 
If instead the request~$?a$ was necessary ({\bf Client2}), then there would be no orchestration for the composition of two {\bf Client2} contracts, because the necessary request is never matched by a corresponding offer. 

 To illustrate the distinction between urgent and lazy necessary transitions, we consider also the server contract shown in Figure~\ref{fig:principals}. If we were to employ the traditional mpc  synthesis, the clients' necessary  requests ($?a$) would be treated as urgent. In such a scenario, the orchestration of the composition between two clients and the server (generated using the mpc synthesis algorithm) would be empty, indicating that no feasible orchestration exists.

However, if the clients' necessary requests ($?a$) are considered lazy instead, an orchestration of the composition between the server and the two clients can be achieved (computed using the orchestration synthesis). This orchestration is depicted in Figure~\ref{fig:orchestration}. In this case, the clients take turns fulfilling their lazy necessary  requests. This alternating behaviour is not possible when the necessary  requests are urgent.

The orchestration in Figure~\ref{fig:orchestration} is obtained after three iterations of the algorithm specified in Definition~\ref{def:synthesisorchestration}. 
Initially, $\mathcal{K}_0 =  \mathcal A = \text{\bf Server}\otimes\text{\bf Client2}\otimes\text{\bf Client2}$ 
and $R_0=Dangling(\mathcal A) = \emptyset$. 

With respect to the orchestration in Figure~\ref{fig:orchestration}, the automaton $\mathcal A$ contains four additional  transitions  that are 
$t_1=[1,0,1] \TRANSS{[-,?a,-]_\Box} [1,1,1]$, 
$t_2=[1,1,0] \TRANSS{[-,-,?a]_\Box} [1,1,1]$, 
$t_3=[1,1,1] \TRANSS{[!\tau,-,-]} [2,1,1]$ and 
$t_4=[2,1,1] \TRANSS{[!a,-,-]} [3,1,1]$.
In the first iteration, $t_1$ and $t_2$ are removed from 
$\mathcal K_1$ because they are request transitions. We have  
$T_{\mathcal{K}_1} = T_{\mathcal{K}_0} \setminus \{t_1,t_2\}$. Since there are no forbidden states, these are the only two transitions that are removed during the first iteration.

Concerning the set of forbidden states $R_1$,  we have that $t_1 \in T^{\Box_l}_{\mathcal A}$ is controllable in $\mathcal K_1$ via  transition $[0,0,0] \TRANSS{[a!,a?,-]_\Box} [1,1,0]$. 
Similarly, $t_2 \in T^{\Box_l}_{\mathcal A}$ is controllable in $\mathcal K_1$ via $[0,0,0] \TRANSS{[a!,-,a?]_\Box} [1,0,1]$. 
Hence, the source states of $t_1$ and $t_2$ will not be added to $R_1$. 
Concerning the set $Dangling(\mathcal K_1)$,  state $[1,1,1]$ was the target of only $t_1$ and $t_2$.   
Moreover,  state $[2,1,1]$ was the target of only  $t_3$. Therefore, states $[1,1,1]$ and $[2,1,1]$ are  unreachable in $\mathcal K_1$. 
We have that 
$R_1=Dangling(\mathcal K_1)=\{[1,1,1],[2,1,1]\}$.
In the subsequent iteration $i=2$, since transition $t_3$ has  target in $R_1$, we have $T_{\mathcal K_2} = T_{\mathcal K_1} \setminus \{t_3\}$, whilst $R_2=R_1$. 

Finally, we reach the fixpoint at iteration $i=3$, where  
$T_{\mathcal K_3} = T_{\mathcal K_2}$ and 
$R_3=R_2$. 
The finalising operations for obtaining the orchestration \text{\bf O} in Figure~\ref{fig:orchestration} from the fixpoint $\mathcal K_3$ consist in removing  the states in $R_3$, i.e., $Q_{\text{\bf O}} = Q_{\mathcal K_3} \setminus R_3$, and removing the remaining unreachable transitions in $\mathcal K_3$. In this case, transition $t_4 \in T_{\mathcal K_3}$ is removed from the orchestration, i.e., $T_{\text{\bf O}} = T_{\mathcal K_3} \setminus \{t_4\}$. 
\end{example}

In the subsequent section, we will delve deeper into additional details and interpretations regarding the semi-controllable transitions of contract automata.

\section{Refined Semi-Controllability}\label{sect:semicontrollability}
We start by introducing a refined notion of semi-controllability to be used in the orchestration synthesis, formalised below.  
After that, we discuss how this refined notion may assist to discard some counter\-intuitive orchestrations.

\begin{definition}[Refined Semi-Controllability]\label{def:refinedsemicontrollability}
Let $\mathcal A$ be an MSCA and let $t = (\vec q_t, \vec a_t, \vec q_t\!') \in T^{\Box_l}_{\mathcal A}$. 
Moreover, given $\mathcal A'\subseteq\mathcal A$, if  
$\exists\,t' = (\vec{q_{t'}}, \vec {a_{t'}}, \vec{q_{t'}}\!') \in T^{\Box_l}_{{\mathcal A}'}$ in $\mathcal A'$ such that the following hold:
\begin{enumerate}
    \item\label{semi:1} 
$\vec a_{t'}$ is a match,  
$\vec q_{t'}, \vec q_{t'}\!' \not\in \textit{Dangling\/}(\mathcal A')$,
$\ithel{\vec{q}_{t}{}}{j} = \ithel{\vec{q}_{t'}{}}{j}$,  $\ithel{\vec{a}_{t}{}}{j} = \ithel{\vec{a}_{t'}{}}{j} = ?a$, for some $j \in 0 \ldots rank(\mathcal A)$;
and 
\item\label{semi:2} there exists a sequence of 
 transitions 
 $t_0, \ldots, t_n$ of $\mathcal A'$ such that 
 $\forall i \in 0\ldots n$,  
$t_i=(\vec{q_i},\vec{a_i},\vec{q_i}')$ and the following hold:
\begin{itemize}
 \item $\vec q_0=\vec q_{t}$; 
 \item $t_n = t'$;
 \item 
$\vec{q_i},\vec{q_i}' \not \in Dangling(\mathcal A ')$; and
\item if $i<n$, then $\ithel{\vec{a_i}}{j}=-$ and  $\vec{q_i}'=\vec{q}_{i+1}$;
\end{itemize}
\end{enumerate}
then $t$ is \emph{controllable\/} in $\mathcal A'$ (via $t'$).
Otherwise, $t$ is \emph{uncontrollable\/} in $\mathcal A'$.
\end{definition}

By comparing Definition~\ref{def:controllabilityorchestration} and 
Definition~\ref{def:refinedsemicontrollability}, we note that only the semi-controllable transitions have  been refined, whilst the others are unaltered. 
Conditions~\ref{semi:1} and~\ref{semi:2} contain the constraints that are used to decide when a semi-controllable transition is controllable or uncontrollable.
The constraints of Condition~\ref{semi:1} are also present in Definition~\ref{def:controllabilityorchestration}. 
The intuition is that a (refined) semi-controllable transition $t$ becomes controllable if (similarly to Definition~\ref{def:controllabilityorchestration}) in a given portion of $\mathcal{A}$, there exists a semi-controllable match transition $t'$, with source and target states not dangling, such that in both $t$ and $t'$ the \emph{same\/} service, in the \emph{same\/} local state, does the \emph{same\/} request. 
Condition~\ref{semi:2} of Definition~\ref{def:refinedsemicontrollability} imposes new further constraints. 
It requires that $t'$ is {\it reachable} from the source state of $t$ through a sequence of transitions where the service performing the request is idle.

\begin{figure}
\centering
\scalebox{.75}{%
\begin{tikzpicture}\node [mybox] (box){%
    \quad\begin{minipage}[t][6cm][c]{\textwidth}
    \begin{tikzpicture}\node [mybox,dashed] (box){%
        \quad\begin{minipage}[t][4cm][c]{.78\textwidth}
        \begin{tikzpicture}\node [mybox,solid] (box){%
            \begin{minipage}[t][2.5cm][c]{.75\textwidth}
		  Hotel~\cite{BBDLFGD20}\\
		  Railway~\cite{BBBC23}\\
            Composition Service~\cite{BB23}\\
            Example~\ref{ex:clients} [Fig.~\ref{fig:orchestration}]\\
            \ldots\hfill{\it\footnotesize has interpretation}
            \end{minipage}
            };
		  \node[fancytitle, right=-60pt, minimum height=16pt] at (box.north east) {\textit{\small Refined}};    
        \end{tikzpicture} 

        \end{minipage}
        };
        \node[fancytitle, right=-45pt, minimum height=16pt] at (box.north east) {\textit{\small ???}};
        \node[right=-78pt,yshift=12pt] at (box.east) %
        {\begin{tabular}{c} {Example~\ref{ex:card}}\\ {[Fig.~\ref{fig:orcCard}]}\end{tabular}};
    \end{tikzpicture}
        \begin{flushright}
            {\it\footnotesize no interpretation?}
        \end{flushright}
    \end{minipage}
    };
    \node[fancytitle, right=-80pt, minimum height=16pt] at (box.north east) {\textit{\small Orchestrations}};
    \node[right=-74pt,yshift=12pt]
    at (box.east) 
    {\begin{tabular}{c} {Example~\ref{ex:alice}}\\ {[Fig.~\ref{fig:orchestrationABC}]}\end{tabular}};
\end{tikzpicture}
}
\caption{A Venn diagram showing the set of orchestrations of contract automata}
\label{fig:venndiagram}
\end{figure}
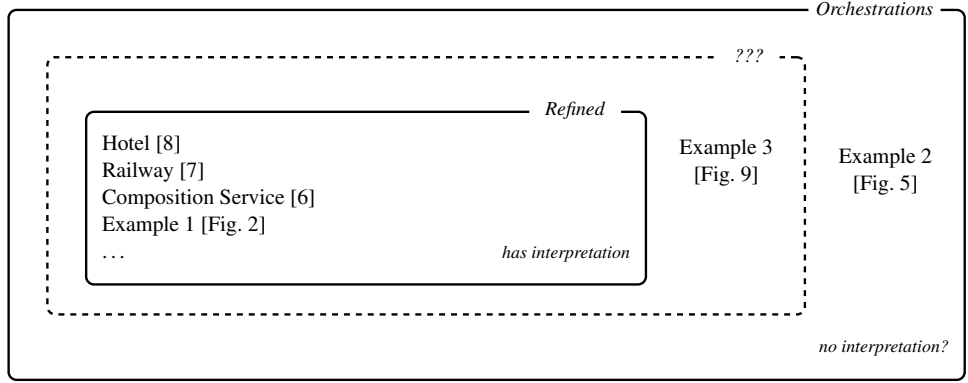

Consider the Venn diagram in Figure~\ref{fig:venndiagram}. 
The outermost set \textit{Orchestrations} contains all orchestrations of contract automata that are computed using the notion of semi-controllability of Definition~\ref{def:controllabilityorchestration}.
The innermost set \textit{Refined} contains only those orchestrations that are computed using the refined notion of semi-controllability in Definition~\ref{def:refinedsemicontrollability}. 
Intuitively, the refined notion imposes a further constraint on \textit{when} a semi-controllable transition is controllable.  As a result, more semi-controllable transitions are uncontrollable than in the previous definition. 
This explains why \textit{Refined} is contained in \textit{Orchestrations}.

All the examples of semi-controllability available in the literature~\cite{BDGDF17,BBDLFGD20,BBP20,BB23} (e.g., Hotel service) and Figure~\ref{fig:orchestration} are orchestrations belonging to the set \textit{Refined} in  Figure~\ref{fig:venndiagram}. 
This means that by updating the notion of semi-controllability, all orchestrations of these examples remain unaltered. 

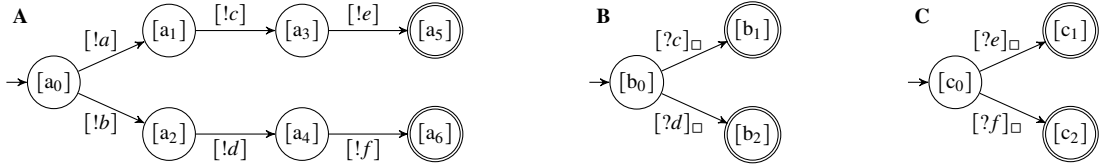
\begin{figure}
\begin{center}
  \begin{tikzpicture}[>=stealth', every state/.style={draw, minimum 
      size=15pt, inner sep=1.5pt}, scale=0.90, node distance=30pt] 
    \node[state] (a0) {\scriptsize $[\text{a}_0]$};
    \node[state,above right=of a0,xshift=8pt,yshift=-16pt] (a1) {\scriptsize $[\text{a}_1]$};
    \node[state,right=of a1] (a3) {\scriptsize $[\text{a}_3]$};
    \node[state,double,right=of a3] (a5) {\scriptsize $[\text{a}_5]$};
    \node[state,below right=of a0,xshift=8pt,yshift=16pt] (a2) {\scriptsize $[\text{a}_2]$};
    \node[state,right=of a2] (a4) {\scriptsize $[\text{a}_4]$};
    \node[state,double,right=of a4] (a6) {\scriptsize $[\text{a}_6]$};
    \path ([xshift=-14pt,yshift=28pt] a0) node {\scriptsize\bf A}; 
    \draw[->] 
    (a0)++(-0.7,0) -- (a0)
    (a0) edge node[xshift=-4pt,yshift=-2pt,above] {\scriptsize $[!a]$} (a1)   
    (a1) edge node[xshift=-2pt,yshift=-2pt,above] {\scriptsize $[!c]$} (a3)
    (a3) edge node[xshift=-2pt,yshift=-2pt,above] {\scriptsize $[!e]$} (a5)
    (a0) edge node[xshift=-4pt,yshift=2pt,below] {\scriptsize $[!b]$} (a2)   
    (a2) edge node[xshift=-2pt,yshift=2pt,below] {\scriptsize $[!d]$} (a4)
    (a4) edge node[xshift=-2pt,yshift=2pt,below] {\scriptsize $[!f]$} (a6); 

    \node[state,right=200pt of a0] (b0) {\scriptsize $[\text{b}_0]$};
    \node[state,double,above right=of b0,xshift=8pt,yshift=-16pt] (b1) {\scriptsize $[\text{b}_1]$};
    \node[state,double,below right=of b0,xshift=8pt,yshift=16pt] (b2) {\scriptsize $[\text{b}_2]$};
    \path ([xshift=-14pt,yshift=28pt] b0) node {\scriptsize\bf B}; 
    \draw[->] 
    (b0)++(-0.7,0) -- (b0)
    (b0) edge node[xshift=-6pt,yshift=-2pt,above] {\scriptsize $[?c]_\Box$} (b1)   
    (b0) edge node[xshift=-6pt,yshift=2pt,below] {\scriptsize $[?d]_\Box$} (b2);
      
    \node[state,right=100pt of b0] (c0) {\scriptsize $[\text{c}_0]$};
    \node[state,double,above right=of c0,xshift=8pt,yshift=-16pt] (c1) {\scriptsize $[\text{c}_1]$};
    \node[state,double,below right=of c0,xshift=8pt,yshift=16pt] (c2) {\scriptsize $[\text{c}_2]$};
    \path ([xshift=-14pt,yshift=28pt] c0) node {\scriptsize\bf C};
    \draw[->] 
    (c0)++(-0.7,0) -- (c0)
    (c0) edge node[xshift=-6pt,yshift=-2pt,above] {\scriptsize $[?e]_\Box$} (c1)   
    (c0) edge node[xshift=-6pt,yshift=2pt,below] {\scriptsize $[?f]_\Box$} (c2);
  \end{tikzpicture} 
\end{center}
\vspace*{-.5\baselineskip}
\caption{\label{fig:principalsABC}Contracts of {\bf A}lice, {\bf B}ob and {\bf C}arl}
\vspace*{.75\baselineskip}
\end{figure}

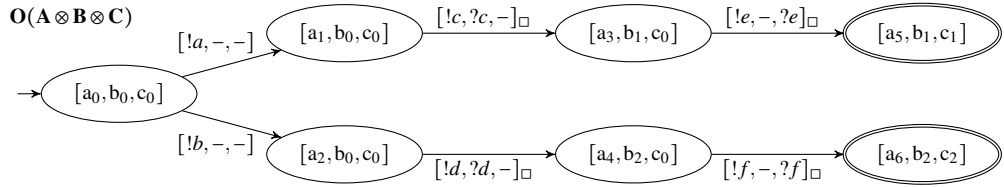
\begin{figure}
\begin{center}
  \begin{tikzpicture}[>=stealth', every state/.style={draw, minimum 
      size=15pt, inner sep=1.5pt}, scale=0.90, node distance=50pt] 
    \node[elliptic state] (a0) {\scriptsize $[\text{a}_0,\text{b}_0,\text{c}_0]$};
    \node[elliptic state,above right=of a0,xshift=8pt,yshift=-28pt] (a1) {\scriptsize $[\text{a}_1,\text{b}_0,\text{c}_0]$};
    \node[elliptic state,right=of a1] (a3) {\scriptsize $[\text{a}_3,\text{b}_1,\text{c}_0]$};
    \node[elliptic state,double,right=of a3] (a5) {\scriptsize $[\text{a}_5,\text{b}_1,\text{c}_1]$};
    \node[elliptic state,below right=of a0,xshift=8pt,yshift=28pt] (a2) {\scriptsize $[\text{a}_2,\text{b}_0,\text{c}_0]$};
    \node[elliptic state,right=of a2] (a4) {\scriptsize $[\text{a}_4,\text{b}_2,\text{c}_0]$};
    \node[elliptic state,double,right=of a4] (a6) {\scriptsize $[\text{a}_6,\text{b}_2,\text{c}_2]$};
    \path ([xshift=-20pt,yshift=32pt] a0) node {\scriptsize${\bf O}(\text{\bf A}\otimes\text{\bf B}\otimes\text{\bf C})$}; 
    \draw[->] 
    (a0)++(-1.5,0) -- (a0)
    (a0) edge node[xshift=-6pt,above] {\scriptsize $[!a,-,-]$} (a1)   
    (a1) edge node[xshift=-2pt,yshift=-2pt,above] {\scriptsize $[!c,?c,-]_\Box$} (a3)
    (a3) edge node[xshift=-2pt,yshift=-2pt,above] {\scriptsize $[!e,-,?e]_\Box$} (a5)
    (a0) edge node[xshift=-6pt,below] {\scriptsize $[!b,-,-]$} (a2)   
    (a2) edge node[xshift=-2pt,yshift=2pt,below] {\scriptsize $[!d,?d,-]_\Box$} (a4)
    (a4) edge node[xshift=-2pt,yshift=2pt,below] {\scriptsize $[!f,-,?f]_\Box$} (a6); 
  \end{tikzpicture} 
\end{center}
\vspace*{-.5\baselineskip}
\caption{\label{fig:orchestrationABC}Orchestration $\text{\bf O}(\text{\bf A}\otimes\text{\bf B}\otimes\text{\bf C})$ of $\text{{\bf A}lice}\otimes\text{{\bf B}ob}\otimes\text{{\bf C}arl}$}
\end{figure}

\begin{example}\label{ex:alice}
We now provide an example of an orchestration belonging to $\textit{Orchestrations} \setminus \textit{Refined}$ (cf.\ Figure~\ref{fig:venndiagram}).
We have three principal contracts, namely {\bf A}lice, {\bf B}ob and {\bf C}arl, depicted in Figure~\ref{fig:principalsABC}.
The contracts of {\bf B}ob and {\bf C}arl perform two alternative necessary requests. 
The contract of {\bf A}lice has two branches. In each branch, a request of {\bf B}ob and a request of {\bf C}arl are fulfilled by corresponding offers.

Using the notion of semi-controllability from Definition~\ref{def:controllabilityorchestration}, the synthesis algorithm of Definition~\ref{def:synthesisorchestration} takes as input the composed automaton and  returns the orchestration of the composition, depicted in Figure~\ref{fig:orchestrationABC}, which is a contract of rank $3$.
Indeed, for each necessary request of each service, there {\it exists} a match transition in the composition where the necessary request is fulfilled by a corresponding offer. 
In other words, for each necessary request of {\bf B}ob and {\bf C}arl, there exists an execution where the request is matched by a corresponding offer. 
For example, the composition $\text{{\bf A}lice}\otimes\text{{\bf B}ob}\otimes\text{{\bf C}arl}$ contains the transition
$t=[\text{a}_1,\text{b}_0,\text{c}_0] \TRANSS{[-,?d,-]_\Box} [\text{a}_1,\text{b}_2,\text{c}_0]$, which is semi-controllable.
According to Definition~\ref{def:controllabilityorchestration}, $t$ is controllable (in $\text{{\bf A}lice}\otimes\text{{\bf B}ob}\otimes\text{{\bf C}arl}$) via $t'=[\text{a}_2,\text{b}_0,\text{c}_0] \TRANSS{[!d,?d,-]_\Box} [\text{a}_4,\text{b}_2,\text{c}_0]$. 
Since $t$ is controllable and it is not in agreement (i.e., the label of $t$ is a request), this transition is pruned during the synthesis of the orchestration. 
We note that $t$ is controllable in $t'$ also in all sub-automaton of the composition computed in the various iterations of the synthesis algorithm, and in the final orchestration depicted in Figure~\ref{fig:orchestrationABC}. 


Using the refined notion of semi-controllability of Definition~\ref{def:refinedsemicontrollability}, the orchestration of $\text{{\bf A}lice} \otimes \text{{\bf B}ob} \otimes \text{{\bf C}arl}$ is empty (i.e., there is no orchestration).  
Consider again transition $t$. 
From state $[\text{a}_1,\text{b}_0,\text{c}_0]$, it is not possible to reach any transition labelled by $[!d,?d,-]_\Box$. It follows that $t$ is uncontrollable. 
Hence, at some iteration $i$ of the orchestration synthesis algorithm in Definition~\ref{def:synthesisorchestration}, state $[\text{a}_1,\text{b}_0,\text{c}_0]$ becomes forbidden and it is added to the set $R_i$. 
At iteration $i+1$, the controllable transition $[\text{a}_0,\text{b}_0,\text{c}_0] \TRANSS{[!a,-,-]_\Box} [\text{a}_1,\text{b}_0,\text{c}_0]$ is 
 pruned because its target state is forbidden. 
At the next iteration ($i+2$), the initial state $[\text{a}_0,\text{b}_0,\text{c}_0]$ becomes forbidden, because there are semi-controllable transitions not in agreement exiting the initial state (e.g., $[\text{a}_0,\text{b}_0,\text{c}_0] \TRANSS{[-,?c,-]_\Box} [\text{a}_0,\text{b}_1,\text{c}_0]$) that are uncontrollable in the sub-automaton whose transitions are $T_{i+2}$. 
Since the initial state is forbidden, it follows that there is no orchestration for $\text{{\bf A}lice} \otimes \text{{\bf B}ob} \otimes \text{{\bf C}arl}$. 

 Indeed,  whenever the state $[\text{a}_1,\text{b}_0,\text{c}_0]$ is reached, although {\bf B}ob and {\bf C}arl are still in their initial state, {\bf B}ob can no longer perform the necessary request~$?d$ and {\bf C}arl can no longer perform the request~$?f$. 
In fact, neither {\bf B}ob nor {\bf C}arl can decide internally which necessary request to execute from their current state.
 For example, there is no trace where the request~$?c$ of {\bf B}ob and the request~$?f$ of {\bf C}arl are matched.
\end{example}

%

The orchestrations belonging to \textit{Refined} (i.e., orchestrations computed using the refined notion of semi-controllability given in Definition~\ref{def:refinedsemicontrollability}) have an intuitive interpretation when compared to the classic notion of uncontrollability. 
We recall that uncontrollable transitions are called {\it urgent} necessary transitions in MSCA, while semi-controllable transitions are called {\it lazy} necessary transitions. 

Intuitively, an urgent transition cannot be delayed, whereas this is the case for a lazy one. 
In a concurrent composition of agents,  the scheduling of concurrent urgent necessary transitions is {\it uncontrollable}. 
Instead, concerning concurrent lazy necessary transitions, each agent {\it internally}  decides its next lazy necessary transition to execute, but the orchestrator schedules when this transition will be executed, i.e., the scheduling is  {\it controllable}.
In Example~\ref{ex:alice}, there is no orchestration because, for example, from state $[\text{a}_1,\text{b}_0,\text{c}_0]$ there is no possible scheduling that allows the services to match all their necessary requests. 
Continuing Example~\ref{ex:clients}, the orchestration in Figure~\ref{fig:orchestration} is non-empty because the scheduling of the actions in the orchestration is {\it controlled} by the orchestrator: one of the two necessary requests is scheduled to be matched only when the server has reached its internal state~$[2]$. 
If instead the clients' necessary request~$?a$ is urgent, then there exists no orchestration of the composition of two clients and the server. 
This is because in this case the scheduling is {\it uncontrollable}: it is not possible to schedule one of the two clients to have its necessary urgent request to be matched only when the server reaches the state~$[2]$. 
In this case, the server should be ready to match the requests whenever they can be executed, without delaying them.


\section{Research Challenges}\label{sect:challenges}
In this section, we describe the currently known limits of the synthesis of orchestrations adopting either Definition~\ref{def:controllabilityorchestration} or Definition~\ref{def:refinedsemicontrollability}, we identify a number of research challenges to overcome these limits, and we propose a research roadmap aimed to tackle these challenges effectively.

First, the notion of semi-controllability introduced in~\cite{BBDLFGD20,BB22OSP} and recalled in 
Definition~\ref{def:controllabilityorchestration} allows to synthesise orchestrations that may sometimes limit the capability of each service to perform internal choices. 
The contract automata formalism abstracts from the way that choices are made. Different implementations are possible in which each service may or may not decide the next step in an orchestration~\cite{BB23}.

Consider again Example~\ref{ex:alice}. Both {\bf B}ob and {\bf C}arl are able to perform two alternative necessary requests from their initial state. 
However, as shown in Figure~\ref{fig:orchestrationABC}, they are forbidden from internally deciding which necessary request is to be executed at runtime.  
If, for example, {\bf B}ob selects the request~$?d$ and {\bf C}arl selects the request~$?e$, then it is not possible for {\bf A}lice to match both requests. 

If we adopt the interpretation given previously (i.e., agents internally choose their necessary transitions and their scheduling is controllable) then we argue that the orchestration computed using Definition~\ref{def:controllabilityorchestration} is too abstract and should in fact be empty. This is indeed the  case if Definition~\ref{def:refinedsemicontrollability} were used instead of Definition~\ref{def:controllabilityorchestration}.

\medskip
\noindent\fbox{%
    \parbox{.985\textwidth}{
The first research challenge is to identify a concrete application of services that perform necessary requests and whose orchestration belongs to the set $\textit{Orchestrations} \setminus \textit{Refined}$. }}
\medskip

Solving this challenge could help provide an intuitive interpretation of these types of orchestrations. 
An application should be identified in which each service statically requires that for each  necessary request there must exist an execution where this is eventually  matched (cf.\ Definition~\ref{def:controllabilityorchestration}). However, during execution, the choice  of which necessary request is to be matched could be  external to the service performing the necessary request. 
Even if the execution of different branches is determined externally, a service contract may still require all branches to be available in the composition. This could be due to the contract's need to enforce certain hyperproperties, such as non-interference or opacity.

Next, we illustrate the second research challenge. 
All examples of orchestrations currently available in the literature~\cite{BBBC23,BB23,BB22OSP,BBP20,BBDLFGD20} reside inside the set \textit{Refined} (cf.\ Figure~\ref{fig:venndiagram}). 
We showed in Example~\ref{ex:alice} an orchestration {\bf O} not belonging to the set \textit{Refined} and we argued that {\bf O} is too abstract and should in fact be empty.
We now provide another example of an orchestration not belonging to the set \textit{Refined}. However, differently from Example~\ref{ex:alice}, in this case the orchestration should not be empty.

\begin{figure}
\begin{center}
  \begin{tikzpicture}[>=stealth', every state/.style={draw, minimum 
      size=15pt, inner sep=1.5pt}, scale=0.90, node distance=30pt] 
    \node[elliptic state] (p0) {\scriptsize $[\text{Dealing}]$};
    \node[state,above right=of p0,xshift=12pt,yshift=-16pt] (p1) {\scriptsize $[\text{P}_1]$};
    \node[state,below right=of p0,xshift=12pt,yshift=16pt] (p2) {\scriptsize $[\text{P}_2]$};
    \node[elliptic state,right=60pt of p0] (p3) {\scriptsize $[\text{Collecting}]$};
    \node[elliptic state,right=90pt of p3] (p4) {\scriptsize $[\text{Card}_2]$};
    \node[elliptic state,double,right=of p4] (p7) {\scriptsize $[\text{Cards}_{21}]$};
    \node[elliptic state,above right=of p3,xshift=16pt,yshift=-16pt] (p5) {\scriptsize $[\text{Card}_3]$};
    \node[elliptic state,below right=of p3,xshift=16pt,yshift=16pt] (p6) {\scriptsize $[\text{Card}_4]$};    
    \node[elliptic state,double,above=of p4] (p8) {\scriptsize $[\text{Cards}_{32}]$};
    \node[elliptic state,double,above=of p7] (p9) {\scriptsize $[\text{Cards}_{31}]$};
    \node[elliptic state,double,below=of p3] (p10) {\scriptsize $[\text{Cards}_{43}]$};
    \node[elliptic state,double,below=of p4] (p11) {\scriptsize $[\text{Cards}_{42}]$};
    \node[elliptic state,double,below=of p7] (p12) {\scriptsize $[\text{Cards}_{41}]$};
    \path ([xshift=-30pt,yshift=68pt] p0) node {\scriptsize\bf Dealer}; 
    \draw[->] 
    (p0)++(-1.5,0) -- (p0)
    (p0) edge node[xshift=-10pt,yshift=-2pt,above] {\scriptsize $[?\textit{pair}_1]$} (p1)
    (p0) edge node[xshift=-10pt,yshift=2pt,below] {\scriptsize $[?\textit{pair}_2]$} (p2) 
    (p1) edge [looseness=1,in=90] node[xshift=-7pt,yshift=-1pt,above] {\scriptsize $[?\textit{pair}_2]$} (p3)
    (p1) edge node[xshift=10pt,yshift=-2pt,above] {\scriptsize $[?\textit{pair}_3]$} (p3)
    (p2) edge node[xshift=10pt,yshift=2pt,below] {\scriptsize $[?\textit{pair}_3]$} (p3)
    (p3) edge node[xshift=30pt,yshift=-2pt,above] {\scriptsize $[!2]$} (p4)
    (p3) edge node[xshift=-2pt,yshift=-2pt,above] {\scriptsize $[!3]$} (p5)
    (p3) edge node[xshift=-2pt,yshift=2pt,below] {\scriptsize $[!4]$} (p6)        
    (p4) edge node[yshift=-2pt,above] {\scriptsize $[!1]$} (p7)
    (p5) edge node[xshift=-4pt,yshift=-2pt,above] {\scriptsize $[!2]$} (p8)
    (p5) edge [out=360,in=210] node[xshift=26pt,yshift=6pt,above] {\scriptsize $[!1]$} (p9)
    (p6) edge node[below] {\scriptsize $[!3]$} (p10)
    (p6) edge node[xshift=-4pt,yshift=2pt,below] {\scriptsize $[!2]$} (p11)
    (p6) edge [out=360,in=150] node[xshift=26pt,yshift=-6pt,below] {\scriptsize $[!1]$} (p12);
  \end{tikzpicture} 
\end{center}
\vspace*{-.5\baselineskip}
\caption{\label{fig:dealer}Contract of the {\bf Dealer}}
\end{figure}
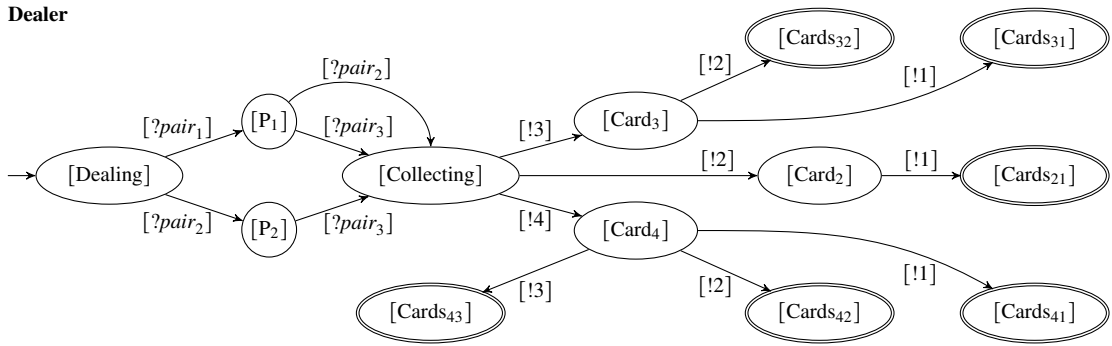

\begin{figure}
\begin{center}
  \begin{tikzpicture}[>=stealth', every state/.style={draw, minimum 
      size=15pt, inner sep=1.5pt}, scale=0.90, node distance=20pt] 
    \node[elliptic state] (p0) {\scriptsize $[\text{Waiting}]$};
    \node[elliptic state,below left=of p0,xshift=-115pt,yshift=-12pt] (p1) {\scriptsize $[\text{Pair}_1]$};
    \node[elliptic state,below=of p0] (p2) {\scriptsize $[\text{Pair}_2]$};    
    \node[elliptic state,below right=of p0,xshift=115pt,yshift=-12pt] (p3) {\scriptsize $[\text{Pair}_3]$};
    \node[elliptic state,double,below left=of p1,xshift=10pt,yshift=-16pt,inner xsep=-.75ex] (p4) {\scriptsize $[\text{Pair}_1\text{Card}_1]$};
    \node[elliptic state,double,below right=of p1,xshift=-10pt,yshift=-16pt,inner xsep=-.75ex] (p5) {\scriptsize $[\text{Pair}_1\text{Card}_3]$};
    \node[elliptic state,double,below left=of p2,xshift=10pt,yshift=-16pt,inner xsep=-.75ex] (p6) {\scriptsize $[\text{Pair}_2\text{Card}_2]$};
    \node[elliptic state,double,below right=of p2,xshift=-10pt,yshift=-16pt,inner xsep=-.75ex] (p7) {\scriptsize $[\text{Pair}_2\text{Card}_4]$};    
    \node[elliptic state,double,below left=of p3,xshift=10pt,yshift=-16pt,inner xsep=-.75ex] (p8) {\scriptsize $[\text{Pair}_3\text{Card}_2]$};
    \node[elliptic state,double,below right=of p3,xshift=-10pt,yshift=-16pt,inner xsep=-.75ex] (p9) {\scriptsize $[\text{Pair}_3\text{Card}_3]$};
    \path ([xshift=-200pt,yshift=18pt] p0) node {\scriptsize\bf Player}; 
    \draw[->] 
    (p0)++(0,.85) -- (p0)
    (p0) edge node[left,xshift=-14pt] {\scriptsize $[!\textit{pair}_1]$} (p1)
    (p0) edge node[right] {\scriptsize $[!\textit{pair}_2]$} (p2) 
    (p0) edge node[right,xshift=14pt] {\scriptsize $[!\textit{pair}_3]$} (p3)
    (p1) edge node[left,xshift=-2pt] {\scriptsize $[?1]_\Box$} (p4)
    (p1) edge node[right,xshift=2pt] {\scriptsize $[?3]_\Box$} (p5)
    (p2) edge node[left,xshift=-2pt] {\scriptsize $[?2]_\Box$} (p6)
    (p2) edge node[right,xshift=2pt] {\scriptsize $[?4]_\Box$} (p7)
    (p3) edge node[left,xshift=-2pt] {\scriptsize $[?2]_\Box$} (p8)
    (p3) edge node[right,xshift=2pt] {\scriptsize $[?3]_\Box$} (p9);
  \end{tikzpicture} 
\end{center}
\vspace*{-.5\baselineskip}
\caption{\label{fig:player}Contract of the {\bf Player}}
\end{figure}
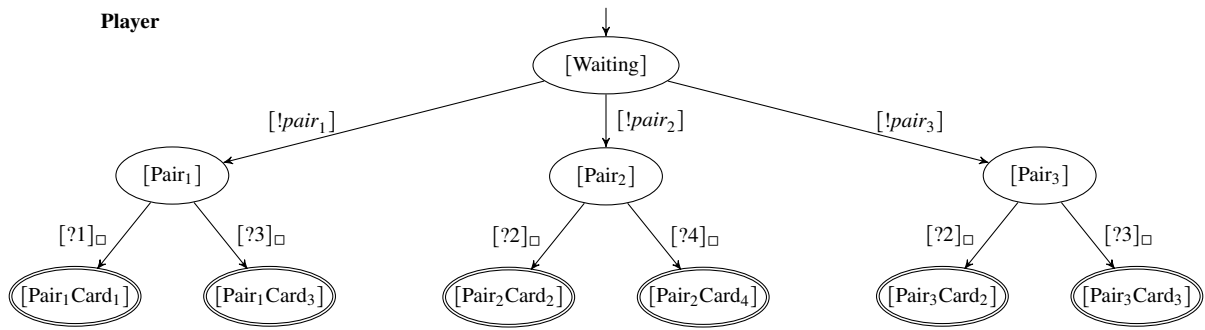

\begin{example}\label{ex:card}
This example involves a simple card game with two players and a dealer. At the beginning of each round, the dealer chooses a pair of cards to deal to each player (i.e., each player receives a pair of cards). 
The dealer can select two out of three different pairs of cards:

\begin{itemize}
\item Pair 1: card 1 and card 3;
\item Pair 2: card 2 and card 4;
\item Pair 3: card 2 and card 3.
\end{itemize}

After the dealer has dealt the pairs of cards, each player selects one of the two cards that was received. Once the players have selected their cards, the dealer collects the selected cards from each player.
The goal of the game is for the dealer to avoid picking up two cards in ascending or equal order, which would result in the dealer losing. In other words, if the dealer picks up a card that is higher than the other card that was picked up or if two cards of the same value are picked up, the dealer loses.
To ensure that the dealer never loses, the dealer has to choose the correct pairs of cards to deal.  There are six possible ways to choose the pairs of cards, but only two of them guarantee a strategy for the dealer to collect the cards selected by the players in descending order.
The strategy for the dealer consists of dealing to the players (in no particular order) Pair~1 and Pair~2. 
Indeed, in the remaining cases there exists the possibility that the players {\it internally} select the same card. 
In this case, there is no way of rearranging the transitions to avoid the same cards being picked by the dealer.

We modelled this above-mentioned problem as an orchestration of contracts, using the refined notion of semi-controllability. 
We only model one round of the game. 
The CA in Figure~\ref{fig:dealer} models the dealer.  
Note that each request can be matched by either of the two players. 
Once the dealer has dealt the pairs of cards, the cards selected by the players are collected. 
Note that the two cards can only be collected in descending order. 
The CA in Figure~\ref{fig:player} models a player. 
Once the player has received a card, the player decides internally which card to select. This internal decision is modelled as a choice among lazy necessary transitions.

The synthesis algorithm adopting the refined notion of semi-controllability from Definition~\ref{def:refinedsemicontrollability} takes as input the composition of the dealer CA and two players CA and returns an empty orchestration. 
To explain why the resulting orchestration is empty, consider Figure~\ref{fig:compositionCard} depicting a portion of the composition of the dealer with two players. 

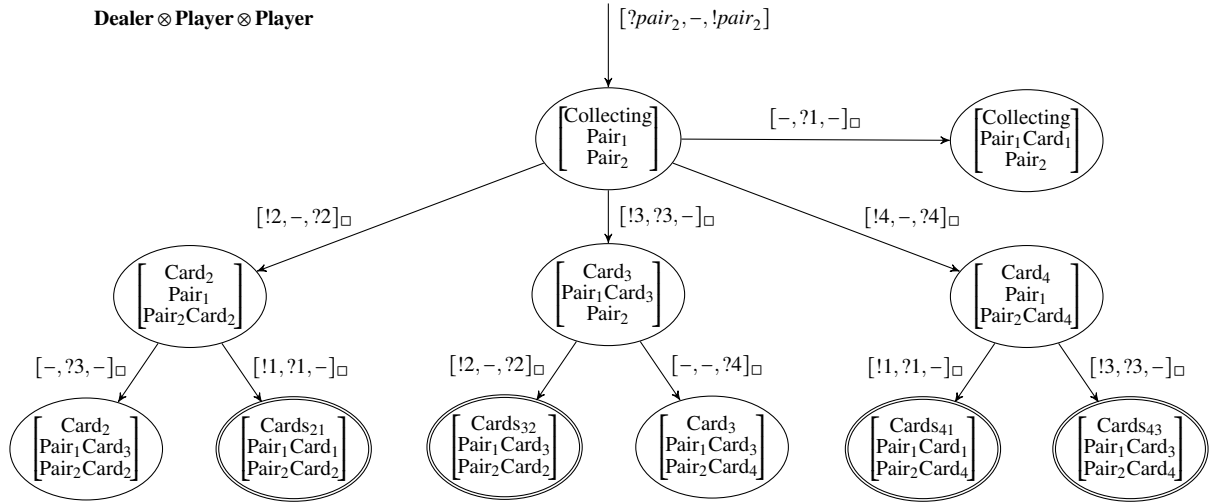
\begin{figure}
\begin{center}\arraycolsep=-1.5pt\def\arraystretch{.85}
  \begin{tikzpicture}[>=stealth', every state/.style={draw, minimum 
      size=15pt, inner sep=1.5pt}, scale=0.90, node distance=20pt] 
    \node[elliptic state,inner xsep=0ex, inner ysep=.25ex] (p0) {\scriptsize $\left[ \begin{array}{c} \text{Collecting} \\ \text{Pair}_1 \\ \text{Pair}_2 \end{array} \right]$};
    \node[elliptic state,below left=of p0,xshift=-104pt,yshift=-18pt,inner xsep=0ex,inner ysep=.25ex] (p1) {\scriptsize $\left[ \begin{array}{c} \text{Card}_2 \\ \text{Pair}_1 \\ \text{Pair}_2\text{Card}_2 \end{array} \right]$};
    \node[elliptic state,below=of p0,inner xsep=0ex,inner ysep=.25ex] (p2) {\scriptsize $\left[ \begin{array}{c} \text{Card}_3 \\ \text{Pair}_1\text{Card}_3 \\ \text{Pair}_2 \end{array} \right]$};    
    \node[elliptic state,below right=of p0,xshift=104pt,yshift=-18pt,inner xsep=0ex,inner ysep=.25ex] (p3) {\scriptsize $\left[ \begin{array}{c} \text{Card}_4 \\ \text{Pair}_1 \\ \text{Pair}_2\text{Card}_4 \end{array} \right]$};
    \node[elliptic state,above=of p3,inner xsep=0ex,inner ysep=.25ex] (p10) {\scriptsize $\left[ \begin{array}{c} \text{Collecting} \\ \text{Pair}_1\text{Card}_1 \\ \text{Pair}_2 \end{array} \right]$};    
    \node[elliptic state,below left=of p1,xshift=16pt,yshift=-16pt,inner xsep=0ex,inner ysep=.25ex] (p4) {\scriptsize $\left[ \begin{array}{c} \text{Card}_2 \\ \text{Pair}_1\text{Card}_3 \\ \text{Pair}_2\text{Card}_2 \end{array} \right]$};
    \node[elliptic state,double,below right=of p1,xshift=-16pt,yshift=-16pt,inner xsep=0ex,inner ysep=.25ex] (p5) {\scriptsize $\left[ \begin{array}{c} \text{Cards}_{21} \\ \text{Pair}_1\text{Card}_1 \\ \text{Pair}_2\text{Card}_2 \end{array} \right]$};
    \node[elliptic state,double,below left=of p2,xshift=16pt,yshift=-16pt,inner xsep=0ex,inner ysep=.25ex] (p6) {\scriptsize $\left[ \begin{array}{c} \text{Cards}_{32} \\ \text{Pair}_1\text{Card}_3 \\ \text{Pair}_2\text{Card}_2 \end{array} \right]$};
    \node[elliptic state,below right=of p2,xshift=-16pt,yshift=-16pt,inner xsep=0ex,inner ysep=.25ex] (p7) {\scriptsize $\left[ \begin{array}{c} \text{Card}_3 \\ \text{Pair}_1\text{Card}_3 \\ \text{Pair}_2\text{Card}_4 \end{array} \right]$};    
    \node[elliptic state,double,below left=of p3,xshift=16pt,yshift=-16pt,inner xsep=0ex,inner ysep=.25ex] (p8) {\scriptsize $\left[ \begin{array}{c} \text{Cards}_{41} \\ \text{Pair}_1\text{Card}_1 \\ \text{Pair}_2\text{Card}_4 \end{array} \right]$};
    \node[elliptic state,double,below right=of p3,xshift=-16pt,yshift=-16pt,inner xsep=0ex,inner ysep=.25ex] (p9) {\scriptsize $\left[ \begin{array}{c} \text{Cards}_{43} \\ \text{Pair}_1\text{Card}_3 \\ \text{Pair}_2\text{Card}_4 \end{array} \right]$};
    \path ([xshift=-170pt,yshift=50pt] p0) node {\scriptsize$\text{\bf Dealer}\otimes\text{\bf Player}\otimes\text{\bf Player}$}; 
    \draw[->] 
    (p0)++(0,2) node[right,yshift=-6pt] {\scriptsize $[?\textit{pair}_2,-,!\textit{pair}_2]$} -- (p0)
    (p0) edge node[left,xshift=-14pt] {\scriptsize $[!2,-,?2]_\Box$} (p1)
    (p0) edge node[right] {\scriptsize $[!3,?3,-]_\Box$} (p2) 
    (p0) edge node[right,xshift=14pt] {\scriptsize $[!4,-,?4]_\Box$} (p3)
    (p0) edge node[above] {\scriptsize $[-,?1,-]_\Box$} (p10) 
    (p1) edge node[left,yshift=2pt] {\scriptsize $[-,?3,-]_\Box$} (p4)
    (p1) edge node[right,yshift=2pt] {\scriptsize $[!1,?1,-]_\Box$} (p5)
    (p2) edge node[left,yshift=2pt] {\scriptsize $[!2,-,?2]_\Box$} (p6)
    (p2) edge node[right,yshift=2pt] {\scriptsize $[-,-,?4]_\Box$} (p7)
    (p3) edge node[left,yshift=2pt] {\scriptsize $[!1,?1,-]_\Box$} (p8)
    (p3) edge node[right,yshift=2pt] {\scriptsize $[!3,?3,-]_\Box$} (p9);
  \end{tikzpicture} 
\end{center}
\vspace*{-.5\baselineskip}
\caption{\label{fig:compositionCard}A fragment of the composition of $\text{\bf Dealer}\otimes\text{\bf Player}\otimes\text{\bf Player}$}
\end{figure}

The state $[\text{Collecting},\text{Pair}_1,\text{Pair}_2]$ is reached when the first player receives $\textit{pair}_1$ and the second player receives $\textit{pair}_2$. 
A symmetric argument holds for state $[\text{Collecting},\text{Pair}_2,\text{Pair}_1]$, not depicted here. 
The transition $$[\text{Card}_2,\text{Pair}_1,\text{Pair}_2\text{Card}_2] \TRANSS{[-,?3,-]_\Box} [\text{Card}_2,\text{Pair}_1\text{Card}_3,\text{Pair}_2\text{Card}_2]$$ is uncontrollable according to Definition~\ref{def:refinedsemicontrollability}. 
Indeed, from state $[\text{Card}_2,\text{Pair}_1,\text{Pair}_2\text{Card}_2]$ it is not possible to reach state $[\text{Collecting},\text{Pair}_1,\text{Pair}_2]$. 
This makes the state $[\text{Card}_2,\text{Pair}_1,\text{Pair}_2\text{Card}_2]$ forbidden. 
Hence, to avoid reaching a forbidden state, the algorithm prunes the transition 
 $$[\text{Collecting},\text{Pair}_1,\text{Pair}_2] \TRANSS{[!2,-,?2]_\Box} [\text{Card}_2,\text{Pair}_1,\text{Pair}_2\text{Card}_2]$$ which is in fact controllable according to Definition~\ref{def:refinedsemicontrollability}. 
 Indeed, from state $[\text{Collecting},\text{Pair}_1,\text{Pair}_2]$ it is possible to reach the transition 
 $$[\text{Card}_3,\text{Pair}_1\text{Card}_3,\text{Pair}_2] \TRANSS{[!2,-,?2]_\Box} [\text{Cards}_{32},\text{Pair}_1\text{Card}_3,\text{Pair}_2\text{Card}_2]$$ via a transition in which the second player is idle. 
However, during the synthesis algorithm also the state $[\text{Card}_3,\text{Pair}_1\text{Card}_3,\text{Pair}_2]$ becomes forbidden due to its outgoing necessary transition, which is uncontrollable according to Definition~\ref{def:controllabilityorchestration}.  
This in turn causes the pruning of transition 
 $$[\text{Collecting},\text{Pair}_1,\text{Pair}_2] \TRANSS{[!3,?3,-]_\Box} [\text{Card}_3,\text{Pair}_1\text{Card}_3,\text{Pair}_2]$$ which is controllable. 
 Once the transition has been pruned, the transition 
 $$[\text{Collecting},\text{Pair}_1,\text{Pair}_2] \TRANSS{[!2,-,?2]_\Box} [\text{Card}_2,\text{Pair}_1,\text{Pair}_2\text{Card}_2]$$ which was previously controllable becomes uncontrollable. 
 This makes the state $[\text{Collecting},\text{Pair}_1,\text{Pair}_2]$ forbidden. 
 Note, however, that $[\text{Collecting},\text{Pair}_1,\text{Pair}_2]$ should {\it not} be forbidden. 
 Indeed, from that state, for each pair of cards selected by the players, the dealer has a strategy to pick them in the correct order:
 \begin{itemize}
    \item if player 1 selects card 1 and player 2 selects card 2, then execute $[!2,-,?2], [!1,?1,-]$;
    \item if player 1 selects card 1 and player 2 selects card 4, then execute $[!4,-,?4], [!1,?1,-]$;
    \item if player 1 selects card 3 and player 2 selects card 2, then execute $[!3,?3,-], [!2,-,?2]$;
    \item if player 1 selects card 3 and player 2 selects card 4, then execute $[!4,-,?4], [!3,?3,-]$.
 \end{itemize}

\begin{figure}
\begin{center}\arraycolsep=-1.5pt\def\arraystretch{.85}
  \begin{tikzpicture}[>=stealth', every state/.style={draw, minimum 
      size=15pt, inner sep=1.5pt}, scale=0.90, node distance=25pt] 
    \node[elliptic state,inner xsep=0ex,inner ysep=1.25ex] (p2) {\scriptsize $\left[ \begin{array}{c} \text{P}_1 \\ \text{Waiting} \\ \text{Pair}_1 \end{array} \right]$};    
    \node[elliptic state,left=of p2,inner xsep=0ex,inner ysep=1.25ex] (p1) {\scriptsize $\left[ \begin{array}{c} \text{P}_1 \\ \text{Pair}_1 \\ \text{Waiting} \end{array} \right]$};
    \node[elliptic state,below=30pt of p1,inner xsep=-.75ex,inner ysep=1.25ex] (p3) {\scriptsize $\left[ \begin{array}{c} \text{Collecting} \\ \text{Pair}_1 \\ \text{Pair}_2 \end{array} \right]$};
    \node[elliptic state,below=30pt of p2,inner xsep=-.75ex,inner ysep=1.25ex] (p4) {\scriptsize $\left[ \begin{array}{c} \text{Collecting} \\ \text{Pair}_2 \\ \text{Pair}_1 \end{array} \right]$};    
    \node[elliptic state,left=38pt of p1,inner xsep=-.75ex,inner ysep=1.25ex] (p5) {\scriptsize $\left[ \begin{array}{c} \text{Card}_4 \\ \text{Pair}_1 \\ \text{Pair}_2\text{Card}_4 \end{array} \right]$};
    \node[elliptic state,left=38pt of p3,inner xsep=-.75ex,inner ysep=1.25ex] (p6) {\scriptsize $\left[ \begin{array}{c} \text{Card}_{3} \\ \text{Pair}_1\text{Card}_3 \\ \text{Pair}_2 \end{array} \right]$};
    \node[elliptic state,below=30pt of p6,inner xsep=-.75ex,inner ysep=1.25ex] (p7) {\scriptsize $\left[ \begin{array}{c} \text{Card}_{2} \\ \text{Pair}_1 \\ \text{Pair}_2\text{Card}_2 \end{array} \right]$};
    \node[elliptic state,double,left=38pt of p7,inner xsep=-.75ex,inner ysep=1.25ex] (p8) {\scriptsize $\left[ \begin{array}{c} \text{Cards}_{21} \\ \text{Pair}_1\text{Card}_1 \\ \text{Pair}_2\text{Card}_2 \end{array} \right]$};    
    \node[elliptic state,double,above=30pt of p8,inner xsep=-.75ex,inner ysep=1.25ex] (p9) {\scriptsize $\left[ \begin{array}{c} \text{Cards}_{32} \\ \text{Pair}_1\text{Card}_3 \\ \text{Pair}_2\text{Card}_2 \end{array} \right]$};
    \node[elliptic state,double,above=30pt of p9,inner xsep=-.75ex,inner ysep=1.25ex] (p10) {\scriptsize $\left[ \begin{array}{c} \text{Cards}_{41} \\ \text{Pair}_1\text{Card}_1 \\ \text{Pair}_2\text{Card}_4 \end{array} \right]$};
    \node[elliptic state,double,above=30pt of p10,inner xsep=-.75ex,inner ysep=1.25ex] (p11) {\scriptsize $\left[ \begin{array}{c} \text{Cards}_{43} \\ \text{Pair}_1\text{Card}_3 \\ \text{Pair}_2\text{Card}_4 \end{array} \right]$};   
    \node[elliptic state,right=160pt of p11,inner xsep=0ex, inner ysep=1.25ex] (p0) {\scriptsize $\left[ \begin{array}{c} \text{Dealing} \\ \text{Waiting} \\ \text{Waiting} \end{array} \right]$}; 
    \node[elliptic state,right=38pt of p2,inner xsep=-.75ex,inner ysep=1.25ex] (p12) {\scriptsize $\left[ \begin{array}{c} \text{Card}_4 \\ \text{Pair}_2\text{Card}_4 \\ \text{Pair}_1 \end{array} \right]$};
    \node[elliptic state,right=38pt of p4,inner xsep=-.75ex,inner ysep=1.25ex] (p13) {\scriptsize $\left[ \begin{array}{c} \text{Card}_{3} \\ \text{Pair}_2 \\ \text{Pair}_1\text{Card}_3 \end{array} \right]$};
    \node[elliptic state,below=30pt of p13,inner xsep=-.75ex,inner ysep=1.25ex] (p14) {\scriptsize $\left[ \begin{array}{c} \text{Card}_{2} \\ \text{Pair}_2\text{Card}_2 \\ \text{Pair}_1 \end{array} \right]$};
    \node[elliptic state,double,right=38pt of p14,inner xsep=-.75ex,inner ysep=1.25ex] (p15) {\scriptsize $\left[ \begin{array}{c} \text{Cards}_{21} \\ \text{Pair}_2\text{Card}_2 \\ \text{Pair}_1\text{Card}_1 \end{array} \right]$};    
    \node[elliptic state,double,above=30pt of p15,inner xsep=-.75ex,inner ysep=1.25ex] (p16) {\scriptsize $\left[ \begin{array}{c} \text{Cards}_{32} \\ \text{Pair}_2\text{Card}_2 \\ \text{Pair}_1\text{Card}_3 \end{array} \right]$};
    \node[elliptic state,double,above=30pt of p16,inner xsep=-.75ex,inner ysep=1.25ex] (p17) {\scriptsize $\left[ \begin{array}{c} \text{Cards}_{41} \\ \text{Pair}_2\text{Card}_4 \\ \text{Pair}_1\text{Card}_1 \end{array} \right]$};
    \node[elliptic state,double,above=30pt of p17,inner xsep=-.75ex,inner ysep=1.25ex] (p18) {\scriptsize $\left[ \begin{array}{c} \text{Cards}_{43} \\ \text{Pair}_2\text{Card}_4 \\ \text{Pair}_1\text{Card}_3 \end{array} \right]$};

    \path ([xshift=-115pt,yshift=25pt] p0) node {\scriptsize $\text{\bf O}(\text{\bf Dealer}\otimes\text{\bf Player}\otimes\text{\bf Player})$}; 
    \draw[->] 
    (p0)++(-1.5,0) -- (p0)
    (p0) edge node[left] {\scriptsize $[?\textit{pair}_1,!\textit{pair}_1,-]$} (p1.north)
    (p0) edge node[right] {\scriptsize $[?\textit{pair}_1,-,!\textit{pair}_1]$} (p2.north) 
    (p1) edge node[right,xshift=-3pt,yshift=6pt] {\scriptsize $[?\textit{pair}_2,-,!\textit{pair}_2]$} (p3)
    (p2) edge node[left,xshift=3pt,yshift=-6pt] {\scriptsize $[?\textit{pair}_2,!\textit{pair}_2,-]$} (p4) 
    (p3) edge node[left] {\scriptsize $[!4,-,?4]_\Box$} (p5)
    (p3) edge node[below] {\scriptsize $[!3,?3,-]_\Box$} (p6)
    (p3) edge node[right,xshift=2pt] {\scriptsize $[!2,-,?2]_\Box$} (p7)
    (p7) edge node[below] {\scriptsize $[!1,?1,-]_\Box$} (p8)
    (p6) edge node[below] {\scriptsize $[!2,-,?2]_\Box$} (p9)
    (p5) edge node[below] {\scriptsize $[!1,?1,-]_\Box$} (p10)
    (p5) edge node[left] {\scriptsize $[!3,?3,-]_\Box$} (p11)    
    (p4) edge node[right,xshift=2pt] {\scriptsize $[!4,?4,-]_\Box$} (p12)
    (p4) edge node[below] {\scriptsize $[!3,-,?3]_\Box$} (p13)
    (p4) edge node[left] {\scriptsize $[!2,?2,-]_\Box$} (p14)
    (p14) edge node[below] {\scriptsize $[!1,-,?1]_\Box$} (p15)
    (p13) edge node[below] {\scriptsize $[!2,?2,-]_\Box$} (p16)
    (p12) edge node[below] {\scriptsize $[!1,-,?1]_\Box$} (p17)
    (p12) edge node[right,xshift=2pt] {\scriptsize $[!3,-,?3]_\Box$} (p18);
  \end{tikzpicture} 
\end{center}
\vspace*{-.5\baselineskip}
\caption{\label{fig:orcCard}Orchestration $\text{\bf O}(\text{\bf Dealer}\otimes\text{\bf Player}\otimes\text{\bf Player})$}
\end{figure}
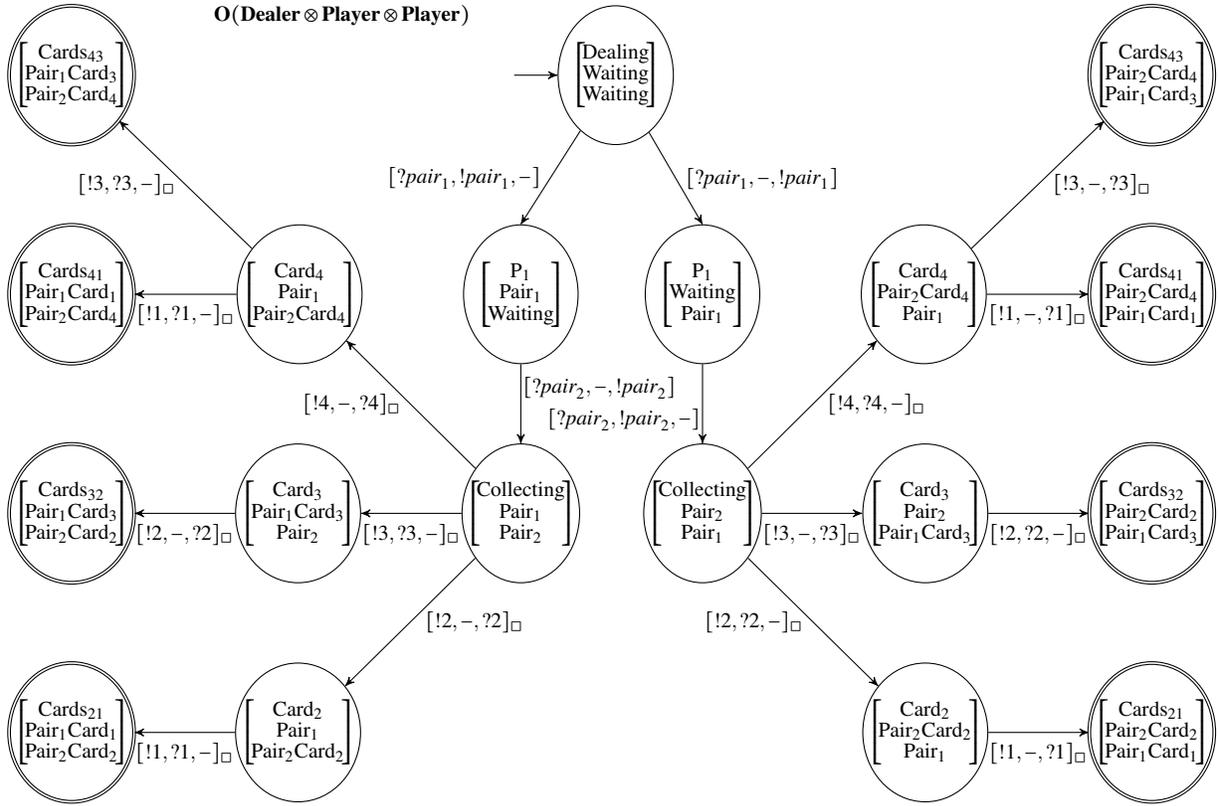

 This example shows that there are cases for which 
 Definition~\ref{def:refinedsemicontrollability} is too restrictive. 
 In this case, the orchestration can be computed using Definition~\ref{def:controllabilityorchestration}, and it is displayed in Figure~\ref{fig:orcCard}.
\end{example}



To better understand the underlying assumption of Definition~\ref{def:refinedsemicontrollability}, we need to decouple the moment in which a service {\it selects} which transition it will execute from the moment in which a service {\it executes} that transition. 
The underlying assumption of Definition~\ref{def:refinedsemicontrollability} is that these two moments are not decoupled. 

For example, the first player whose internal state is $\text{Pair}_1$ could select and execute~$?3$ also from state $[\text{Card}_2,\text{Pair}_1,\text{Pair}_2\text{Card}_2]$, while the strategy described above assumes that the player selects a card in state $[\text{Collecting},\text{Pair}_1,\text{Pair}_2]$.
In fact, the current implementation of the contract automata runtime environment {\tt CARE}~\cite{BB23} allows the decoupling of these two moments. Once state $[\text{Collecting},\text{Pair}_1,\text{Pair}_2]$ is reached, the orchestrator interacts with both players and, based on their choices, correctly schedules the transitions of the dealer and the players. 
This means that the players select their next action in state $[\text{Collecting},\text{Pair}_1,\text{Pair}_2]$ and afterwards their execution is bounded to the transition they have selected. 
 Summarising, Example~\ref{ex:alice} has showed that in some cases Definition~\ref{def:controllabilityorchestration} is too abstract, whereas Example~\ref{ex:card} has showed that in some cases Definition~\ref{def:refinedsemicontrollability} is too restrictive. 
 
\medskip
\noindent\fbox{%
    \parbox{.985\textwidth}{
 The second research challenge is to identify a notion of semi-controllability capable of discarding orchestrations such as the one in Example~\ref{ex:alice} and providing non-empty orchestrations in scenarios such as the one described in Example~\ref{ex:card}. 
 }}
 \medskip

 The resulting, currently unknown set of orchestrations that would be identified by the notion of semi-controllability that solves this challenge is depicted in Figure~\ref{fig:venndiagram} with dashed lines. 

We continue by discussing further research challenges for the orchestration synthesis of contract automata. An important aspect is the ability to scale to large orchestrations when many service contracts are composed. 
We note that computing Definition~\ref{def:refinedsemicontrollability} is harder than computing Definition~\ref{def:controllabilityorchestration}, due to the additional constraint of reachability which requires a visit of the automaton. 
Decoupling the moment in which a service selects a choice from the moment in which the selected choice is executed, could further increase the hardness of deciding when a lazy necessary transition is controllable. 

Consider again the CA in Figure~\ref{fig:principals}. From their initial state, both {\bf B}ob and {\bf C}arl have two choices. If, instead of two principals, we had ten principals whose behaviour is similar to that of {\bf B}ob and {\bf C}arl, then there would be $2^{10}$ possible combinations of (internal) choices the services could make.

\medskip
 \noindent\fbox{%
    \parbox{.985\textwidth}{
    The third research challenge is to provide scalable solutions for synthesising orchestrations. 
    }}
\medskip

Generally speaking, the behaviour of an orchestration that belongs to the unknown dotted set  of Figure~\ref{fig:venndiagram} must be a sub-automaton of an orchestration computed using Definition~\ref{def:controllabilityorchestration} and a super-automaton of an  orchestration computed using Definition~\ref{def:refinedsemicontrollability}.
Indeed, Definition~\ref{def:controllabilityorchestration} can be used as an upper bound and Definition~\ref{def:refinedsemicontrollability} as a lower bound to approximate the behaviour of such an orchestration. 
    
Finally, we discuss the last research challenge identified in this paper. 
We previously formalised the notion of lazy necessary request that is semi-controllable according to either Definition~\ref{def:controllabilityorchestration} or 
Definition~\ref{def:refinedsemicontrollability}. 
We noted that Definition~\ref{def:controllabilityorchestration}  may exclude the case in which, in the presence  of a choice, a service may internally select its necessary transition.
Instead, Definition~\ref{def:refinedsemicontrollability} may exclude the case in which, in the presence of a choice, the moment in which the service internally selects its necessary transition is decoupled from the moment in which the selected necessary transition is executed. 
In other words, we identified two {\it requirements} that an orchestration of services should satisfy: {\it independence} and {\it decoupling} of choices. 

\medskip
 \noindent\fbox{%
    \parbox{.985\textwidth}{
The fourth research challenge is to consolidate a set of requirements that a desirable orchestration of service contracts  must satisfy.
    }}
\medskip

The requirements that would solve this challenge should be established incrementally, as discussed in this paper. 
Formal definitions of necessary service transitions and  practical examples are useful to identify the ideal set of requirements that an orchestration of services should satisfy. 
Of course, these requirements are entangled with the underlying execution support of an orchestration of services, which was recently proposed in~\cite{BB23}.

\subsection{Research Roadmap}

We have presented a series of research challenges associated with the orchestration of contract automata. 
We now propose a potential research roadmap  aimed at tackling these  challenges effectively. However, it is necessary to further examine the concepts described below to determine their validity.

\paragraph{Specifying Choices}

We propose to concretise the selection of the next transition to execute at contract automata level, distinguishing between internal and external selections. 
Presently, this distinction is abstracted away within contracts and handled by the underlying execution support. 
Our rationale is that abstracting from the selection process may lead to scalability challenges. 
Specifically, if a transition is selected internally, it must always be available, whereas an externally selected transition can be removed from the orchestration. 
In essence, internal selection imposes stricter requirements than external selection. 
Consequently, treating all selections as internal to ensure independence of choice leads to larger state spaces. 
For instance, the issue highlighted in Example~\ref{ex:alice} arises due to the presence of externally selected transitions. By allowing contracts to specify which transitions are internally or externally selected, we can potentially reduce the state space, as compared to considering all choices as internal.

Pursuing the above has important implications. 
Firstly, it necessitates updating accordingly the underlying execution support, {\tt CARE}, to align it with the contract automata specifications. 
This entails reducing the implementation freedom for each choice to adhere to the contract's explicit selection of the next transition. 
By explicating choices within contracts, we establish the interpretation of necessary requests discussed in this paper. 
In this interpretation, a service {\it internally} decides to perform a necessary request,  but the scheduling of the execution of the request is controlled by the orchestrator. 
In other words, optional actions are externally selected, whereas necessary actions are internally selected. 
Consequently, by explicitly stating choices in contracts, we can address the first research challenge. 
Indeed, all necessary requests would be internally selected.  Scenarios like the one outlined in the context of the first research challenge (i.e., external necessary requests) would be practically ruled out. 


Another implication relates to the fourth research challenge, which entails consolidating a set of requirements for effective orchestrations. Notably, if choices are explicitly specified in contracts, the requirement of independence of choice can be removed.

\paragraph{Implementing the Decoupling of Choices} 
The second research challenge, as mentioned previously, revolves around the absence of the decoupling of choice requirement in both Definitions~\ref{def:controllabilityorchestration} and~\ref{def:refinedsemicontrollability}. 
This requirement suggests a potential implementation of semi-controllable transitions and may help identify the currently unknown set of orchestrations in Figure~\ref{fig:venndiagram}.
Currently, a semi-controllable transition is defined as a transition that can be either controllable or uncontrollable based on a global condition of the automaton. 
However, decoupling the moment when a service internally selects a transition from the moment when the transition is executed 
might require splitting a semi-controllable transition into two distinct transitions. 

Reasoning in this way suggests that a semi-controllable transition could potentially be represented as two consecutive transitions. 
 The first transition would be  uncontrollable,  capturing the internal selection, while the subsequent transition would be  controllable and responsible for executing the action. 
For example, consider a semi-controllable transition ${\tt [q]} \TRANSS{ [?a]_{\Box l}} {\tt [q']}$, which would be split into two transitions: $t_1={\tt [q]} \TRANSS{[\tau]_{\Box u}} {\tt [i]}$ and $t_2={\tt [i]} \TRANSS{[?a]} {\tt [q']}$.
Here, $t_1$ represents an uncontrollable silent transition to an intermediate, non-final state, while $t_2$ is controllable and executes the action. 
This approach suggests  that the orchestrator cannot control the internal selection made with $t_1$, but it can control and schedule the execution of the action indicated by $t_2$. Moreover, an important consequence of the fact that the intermediate state is non-final, is that $t_2$ must eventually be executed.

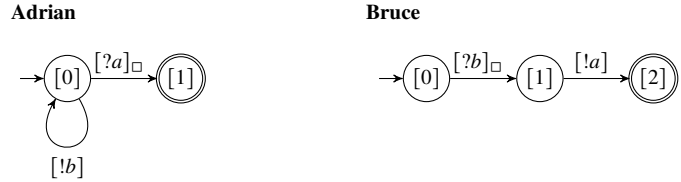
\begin{figure}
\begin{center}
  \begin{tikzpicture}[>=stealth', every state/.style={draw, minimum 
      size=15pt, inner sep=1.5pt}, scale=0.90, node distance=25pt] 
    \node[state] (0) {\scriptsize $[0]$};
    \node[state,double,right=of 0] (1) {\scriptsize $[1]$};
    \path ([xshift=-10pt,yshift=28pt] 0) node {\scriptsize\bf Adrian}; 
    \draw[->] 
    (0)++(-0.7,0) -- (0)
    (0) edge [loop below,out=300,in=240,looseness=8] node[below] {\scriptsize $[!b]$} (0)   
    (0) edge node[xshift=-2pt,yshift=-2pt,above] {\scriptsize $[?a]_\Box$} (1); 

    \node[state,right=75pt of 1] (0bis) {\scriptsize $[0]$};
    \node[state,right=of 0bis] (1bis) {\scriptsize $[1]$};
    \node[state,double,right=of 1bis] (2bis) {\scriptsize $[2]$};
    
    \path ([xshift=-14pt,yshift=28pt] 0bis) node {\scriptsize\bf Bruce}; 
    \draw[->] 
    (0bis)++(-0.7,0) -- (0bis)   
    (0bis) edge node[xshift=-2pt,yshift=-2pt,above] {\scriptsize $[?b]_\Box$} (1bis)
    (1bis) edge node[xshift=-2pt,yshift=-2pt,above] {\scriptsize $[!a]$} (2bis);     
  \end{tikzpicture} 
\end{center}
    \vspace*{-.5\baselineskip}
    \caption{\label{fig:loop}Two contracts whose orchestration requires further investigation}
\end{figure}

Further exploration is required to determine whether this interpretation of semi-controllability solves the third research challenge. In particular, there are still corner cases that require further investigation. For instance, consider the contracts in Figure~\ref{fig:loop}. 
Although an orchestration could be obtained by matching the necessary request {\tt ?b} of Bruce first and only afterwards the necessary request {\tt ?a} of Adrian, this orchestration is not supported by the notion of semi-controllability outlined above.
In this orchestration, Adrian internally selects the request {\tt ?a} and the orchestrator schedules the request of Adrian to be matched later after Adrian matches the request of Bruce.

Furthermore, we envision the establishment of a clear separation between optional and necessary transitions on the one hand and controllable and uncontrollable transitions on the other. 
All necessary requests should be categorised as lazy/semi-controllable, thus effectively excluding urgent necessary requests from contracts. 
This implies that contract automata with optional and necessary transitions should be transformed into automata with solely controllable and uncontrollable transitions, which are known as plant automata in supervisory control theory. It is worth noting that all uncontrollable transitions will serve as silent moves to represent the internal selection of a necessary transition.

\paragraph{Experimental Validation of Performance}

The third research challenge highlights the issue of scalability and proposes the adoption of Definition~\ref{def:controllabilityorchestration} as an upper bound for the set of orchestrations. However, it remains unclear whether the synthesis process using Definition~\ref{def:controllabilityorchestration} is faster compared to synthesising using the mapped plant automaton as suggested earlier. Definition~\ref{def:controllabilityorchestration} necessitates a visit of the automaton at each iteration of the 
synthesis process to determine whether a semi-controllable transition is controllable or uncontrollable. This requirement is not present in a plant automaton consisting solely of controllable and uncontrollable transitions.
On the other hand, the suggested mapping approach increases the state space of the automata by introducing an additional state for each necessary transition. As a result, it is essential to conduct further experimental research to assess the effectiveness of utilising Definition~\ref{def:controllabilityorchestration} as an upper bound for the set of orchestrations. This research should involve measuring the performance and efficiency of the synthesis process when employing Definition~\ref{def:controllabilityorchestration} and comparing it with the approach based on the mapped plant automaton.


\section{Conclusion}\label{sect:conclusion}
We have presented a number of research challenges related to the orchestration synthesis of contract automata. Initially, we proposed a novel refined definition of semi-controllability and compared it to the current definition through illustrative examples. 
We identified various sets of orchestrations, as showed in Figure~\ref{fig:venndiagram}. Additionally, we informally discussed two prerequisites that the orchestration of contracts should satisfy: independence and decoupling of choices. Furthermore, we evaluated the current formal definitions of semi-controllability based on these requirements, which generated a series of research questions regarding the orchestration synthesis of contract automata, to be addressed in future work, possibly by following the proposed research roadmap.

\paragraph{Acknowledgements} We would like to thank the reviewers for their useful comments. 
This work has been partially supported by  the Italian MIUR PRIN 2017FTXR7S project IT MaTTerS (Methods and Tools for Trustworthy Smart Systems) and the Italian MUR PRIN 2020TL3X8X project T-LADIES (Typeful Language Adaptation for Dynamic, Interacting and Evolving Systems).

\bibliographystyle{eptcs}
\bibliography{bib}

\end{document}